\documentclass[fleqn,usenatbib]{rasti}

\usepackage{newtxtext,newtxmath}

\usepackage[T1]{fontenc}

\DeclareRobustCommand{\VAN}[3]{#2}
\let\VANthebibliography\thebibliography
\def\thebibliography{\DeclareRobustCommand{\VAN}[3]{##3}\VANthebibliography}

\usepackage{graphicx}	
\usepackage{amsmath}	

\usepackage{tikz}

\usetikzlibrary{arrows.meta}
\usepackage{booktabs} 
\usepackage{threeparttable}
\usepackage{multirow} 
\usepackage{array} 
\usepackage{tabularx}
\usepackage{subcaption}
\usepackage{comment}

\title[The Spectrometer Development of CosmoCube to Detect 21-cm Signal]{The Spectrometer Development of CosmoCube, Lunar Orbiting Satellite to Detect 21-cm Hydrogen Signal from Cosmic Dark Ages}

\author[K. Artuc et al.]{
Kaan Artuc,$^{1,2}$\thanks{E-mail: ka560@cam.ac.uk}
and Eloy de Lera Acedo$^{1,2}$
\\
$^{1}$Cavendish Laboratory, University of Cambridge, J. J. Thomson Avenue, Cambridge, CB3 0HE, UK\\
$^{2}$Kavli Institute for Cosmology, Madingley Road, Cambridge, CB3 0HA, UK
}

\date{Accepted XXX. Received YYY; in original form ZZZ}

\pubyear{2022}

\begin{document}
\label{firstpage}
\pagerange{\pageref{firstpage}--\pageref{lastpage}}
\maketitle

\begin{abstract}
The cosmic Dark Ages represent a pivotal epoch in the evolution of the Universe, marked by the emergence of the first cosmic structures under the influence of dark matter. The 21-cm hydrogen line, emanating from the hyperfine transition of neutral hydrogen, serves as a critical probe into this era. We describe the development and implementation of the spectrometer for CosmoCube, a novel lunar orbiting CubeSat designed to detect the redshifted 21-cm signal within the redshift range of 13 to 150. Our instrumentation utilizes a Xilinx Radio Frequency System-on-Chip (RFSoC), which integrates both Analog-to-Digital Converters (ADCs) and Digital-to-Analog Converters (DACs), tailored for the spectrometer component of the radiometer. This system is characterized by a 4096 FFT length at 62.5 kHz steps using a Polyphase Filter Bank (PFB), achieving an average Effective Number of Bits (ENOB) of 11.5 bits throughout the frequency of interest, from 10 MHz to 100 MHz. The spectrometer design is further refined through loopback tests involving both DAC and ADC of the RFSoC, with DAC outputs varying between high (+1 dBm) and low (-3 dBm) power modes to characterize system performance. The power consumption was optimized to 5.45 W using three ADCs and one DAC for the radiometer. Additionally, the stability of the ADC noise floor was investigated in a thermal chamber with environmental temperatures ranging from 5°C to 40°C. A consistent noise floor of approximately -152.5 dBFS/Hz was measured, with a variation of $\pm$0.2 dB, ensuring robust performance under varying thermal conditions.
\end{abstract}




\section{Introduction}

During the Dark Ages, following the Universe's cooling to a point where atomic formation became possible, a shift occurred from an intensely hot and luminous environment to a cold and dark state. Over time, gravitational forces gradually amplified small perturbations in the distribution of gas, leading to the emergence of expansive voids and substantial hydrogen clouds \citep{Peebles1981}. The 21-cm line, emitted through the hyperfine transition of neutral hydrogen, allows for the observation of how intervening hydrogen affects Cosmic Microwave Background (CMB) radiation at redshifted 21 cm wavelengths. Depending on whether the spin temperature of hydrogen is lower or higher than the CMB temperature, the 21-cm signal can manifest as slight absorption or emission, respectively, facilitating a detailed exploration of various epochs of the Universe. Observation of the 21-cm signal allows us to trace the distribution and density fluctuations of hydrogen clouds influenced by the underlying dark matter framework \citep{Morales2010}. This methodology provides insights into the gravitational forces necessary for the aggregation of these clouds into proto-stellar bodies and contributes to our understanding of cosmic structure formation \citep{Barkana2016}. By mapping variations in the 21-cm signal across different regions in the Dark Ages, we can infer the density and distribution of dark matter, shedding light on its role in catalyzing the birth of the first stars and galaxies \citep{Pritchard2012}. Focusing on 21-cm cosmology not only enhances our understanding of dark matter's influence on early star formation but also deepens our comprehension of the Universe's evolutionary history during its most formative stages \citep{Cohen2017}.

The 21-cm hydrogen signal has been leveraged to provide constraints on cosmological parameters, contributing to our understanding of the Universe's physical constants \citep{reis2020}. \citet{Chen2004} explored the impact of dark matter, dark energy, and baryonic physics on the 21-cm signal through simulations and theoretical analyses, highlighting its dependence on fundamental properties. \citet{Furlanetto2006} investigated the interplay of various processes during the Dark Ages and reionization, advancing our comprehension of the 21-cm signal's evolution and enabling the estimation of cosmological parameters.

\definecolor{nicegreen}{RGB}{0,158,115} 
\begin{figure*}
  \centering
  \begin{tikzpicture}
    \node[anchor=south west,inner sep=0] (image) at (0,0) {\includegraphics[width=\textwidth]{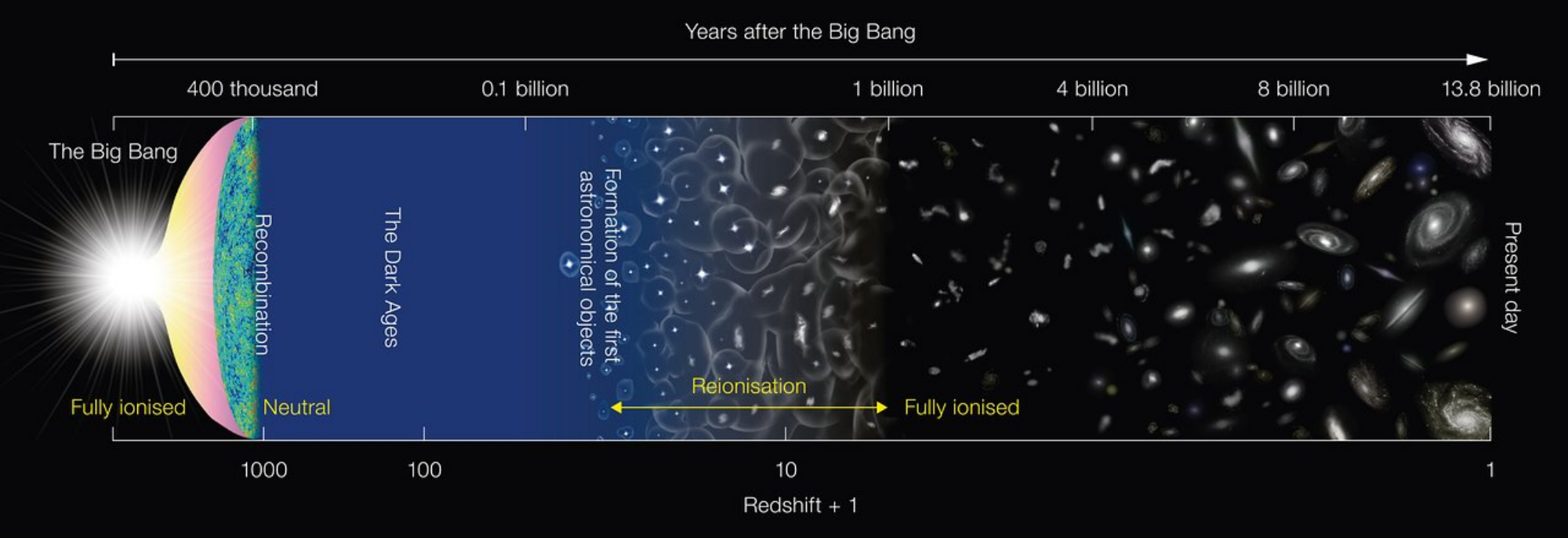}};
    \begin{scope}[x={(image.south east)},y={(image.north west)}]
        \shade[left color=transparent,right color=transparent,middle color=nicegreen,opacity=0.4] (0.20,0.25) rectangle (0.46,0.34);
        \node at (0.33,0.295) [text=white] {CosmoCube};
    \end{scope}
  \end{tikzpicture}
  \caption{This illustration outlines the epochs from the Big Bang to the present day. The CosmoCube aims to probe the faint 21-cm line from the 'Dark Ages'—the period before the first stars ignited—through to the 'Reionization' phase when the Universe's neutral hydrogen was ionized by the first luminous structures. By capturing this faint signal, the CosmoCube seeks to shed light on the structure and evolution of the early Universe.  Ground experiments for the 21-cm signal experiment start where CosmoCube's coverage ends. (Credit: ESO)}
  \label{fig:CosmoCube coverage}               
\end{figure*}

Observing the 21-cm signal faces significant challenges, primarily due to foreground contamination by galactic synchrotron radiation, extragalactic sources, and free-free emissions, which significantly overpower the target signal \citep{Bernardi2010, Ghosh2012, Offringa2013, Barry2016, Lian2020}. Ground-based observations, in particular, grapple with additional complications from the Earth's ionosphere and terrestrial radio frequency interference (RFI), which distort and obscure the signal further \citep{Lawrence1964, Vedantham2013, Datta2016, Shen2021, Shen2022}. Detecting the low-frequency 21-cm signal from the Dark Ages from Earth is particularly challenging due to the Earth's ionosphere, which introduces substantial distortions and absorption effects at these frequencies. The ionosphere's electron density fluctuations can cause signal propagation delays and scattering, distorting the 21-cm signal. Ground spillover, caused by the scattering and reflection of radio waves from the Earth's surface, worsens the challenge, contributing to the overall noise floor and interfering with the cosmological signal, thereby limiting the sensitivity and precision of Earth-conducted observations. The strategic placement of a ground plane below the antenna is a common mitigation technique to address this issue \citep{Singh2018, Mahesh2021, Cumner2022}.

To address these challenges, calibration techniques and technological innovations have been developed through quantification of these phenomenons \citep{Byrne2019, Shen2021}. Despite the efforts of developing robust radiometers, the persistent issues of ionospheric distortion and terrestrial RFI underline the inherent limitations of ground-based observations \citep{Datta2016, Offringa2013}. This realization has spurred interest in alternative observation strategies, including the potential for space-based platforms. A pioneering approach involves deploying a satellite to orbit around the Moon, leveraging the Moon's far side as a shield against Earth's RFI and ionospheric effects, offering a clearer window to the cosmic 21-cm signal and a step forward in unraveling the mysteries of the Universe's early epochs.

Earth-based low-frequency global signal experiments such as EDGES \citep{Bowman2008}, SARAS \citep{Singh2018}, LEDA \citep{Bernardi2016}, PRIZM \citep{Philip2019}, MIST \citep{Monsalve2023}, and REACH \citep{Acedo2022} aim to constrain the physics of the Cosmic Dawn (CD) and the Epoch of Reionization (EoR). In contrast, earth-based interferometer radio experiments like LOFAR \citep{Haarlem2013}, HERA \citep{DeBoer2017}, MWA \citep{Tingay2013}, and the future SKA Observatory \citep{ska2015} are exploring the 21-cm signal to understand the early Universe's spatial power spectrum. The EDGES experiment \citep{Bowman2008}, notable for its claim of detecting the global 21-cm signal at 78 MHz using two low-band dipole antennae operating between 50 and 100 MHz, has ignited a debate over its interpretation. This claim has been met with skepticism by other research groups amid concerns on the data analysis used \citep{hills2018, sims2019, bevins2022, singh2019}.

Research into detecting the 21-cm signal from the Dark Ages is expanding, with missions such as LuSEE-Night \citep{bale2023}, ROLSES \citep{burns2021}, and LCRT \citep{goel2022probing} focusing on surface single antenna observations, surface interferometry missions like FarView \citep{Polidan2024}, ALO \citep{Wolt2024}, FARSIDE \citep{burns2021lunar}, and LARAF \citep{chen2024LARAF}, and interferometry in orbit such as DSL \citep{Chen2021}, along with single antenna in orbit missions like NCLE \citep{BENTUM2020856}, PRATUSH \citep{pratush2024}, and this work, CosmoCube, all concentrating on lunar-based observations. LuSEE-Night plans to utilize a lander with monopole antennas on the Moon's far side, aiming for frequencies between 0.5 and 50 MHz. ROLSES has successfully landed, but the lander tipped over and ended up at a 30-degree angle to the horizontal \citep{nasaROLSES}. While the four STACER antennas—spring-tensioned, collapsible antennas designed for deployment in space—were deployed, their exact final orientation remains uncertain. Nevertheless, the investigators confirm that data have been transmitted from the instrument and are now under analysis. The Chang’e 4 mission \citep{jia2018} also contributes with its rover equipped with antennas optimized for 40-80 MHz for one of its channels, demonstrating a trend towards more compact and efficient designs for lunar and space-based low-frequency signal detection \citep{lai2020}. Proposed FarView, a lunar radio array mission, utilizes 100,000 dipole antennas, manufactured in situ on the Moon using metals extracted from the lunar regolith. Operating across a frequency range of 5 to 50 MHz, the array employs interferometry to enhance the resolution and depth of cosmic observations.

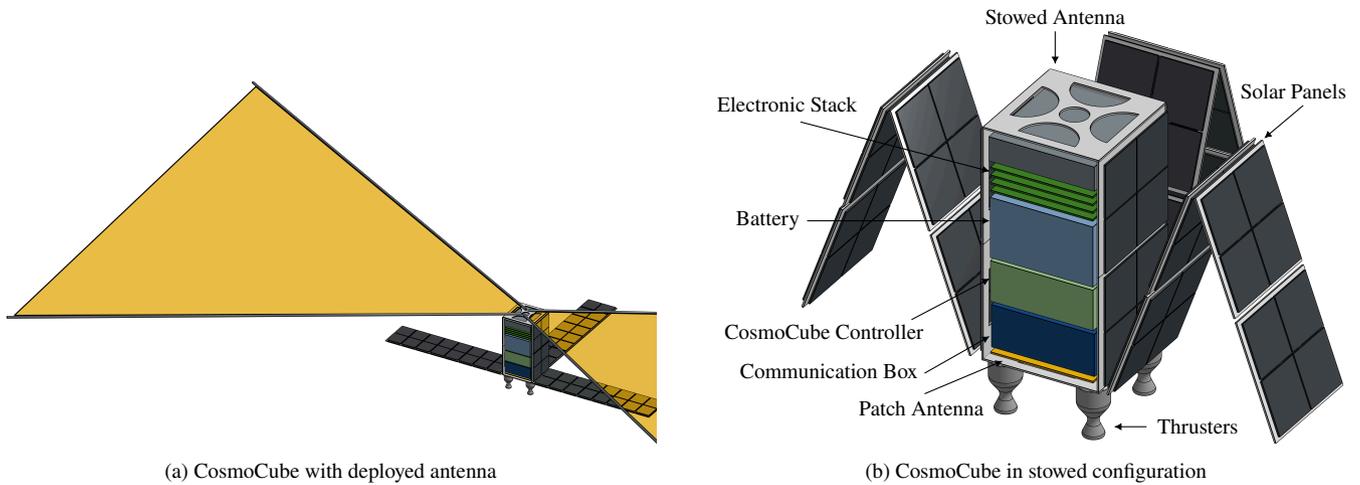
\begin{figure*}
    \centering
    \begin{subfigure}[t]{\columnwidth}
        \centering
        \includegraphics[width=\textwidth, trim=0cm 12cm 0cm 6.4cm, clip]{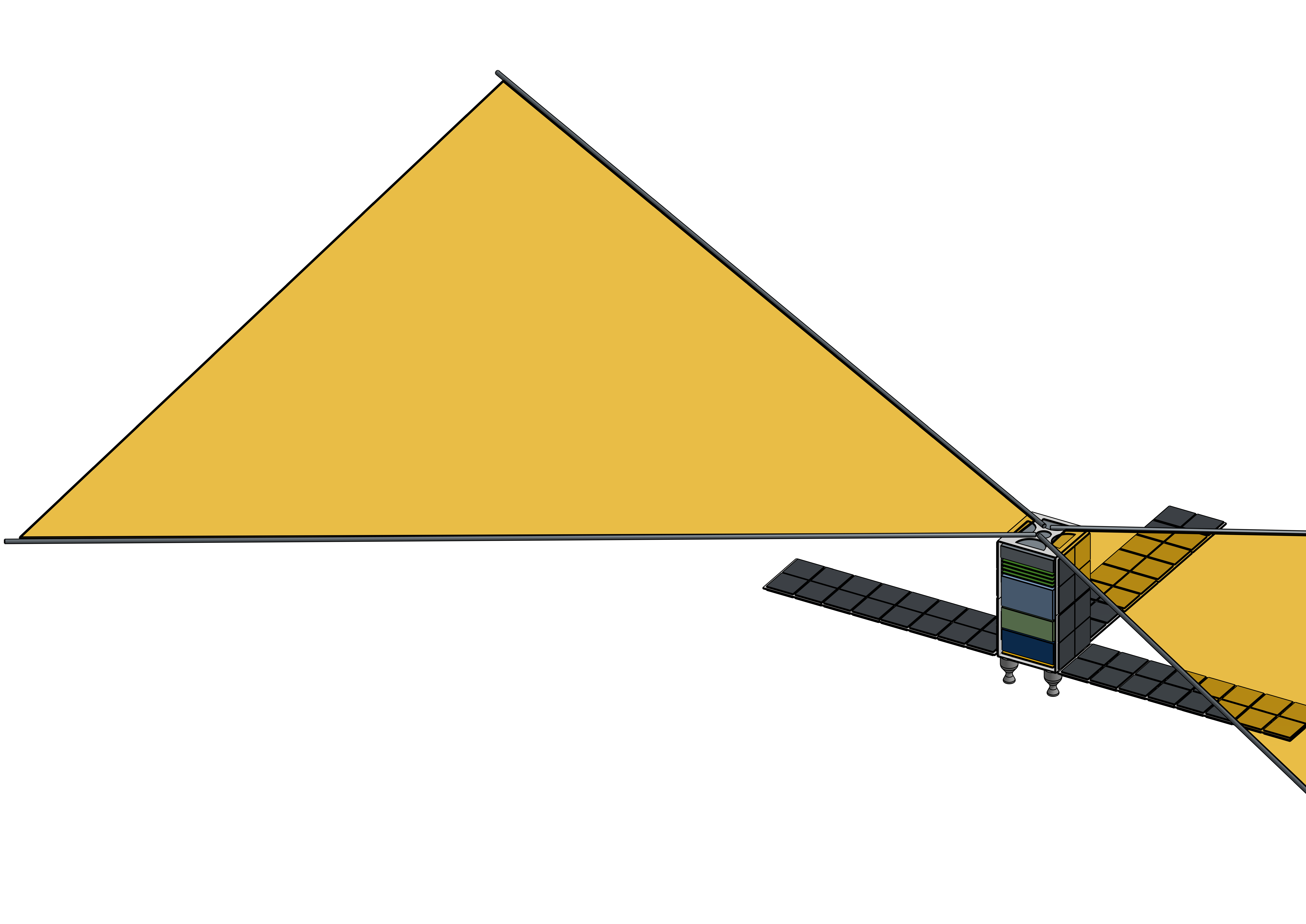}
        \caption{CosmoCube with deployed antenna}
        \label{fig:cosmocube_annotated_a}
    \end{subfigure}%
    \hfill
    \begin{subfigure}[t]{\columnwidth}
        \centering
        \begin{tikzpicture}
            \node[anchor=south west,inner sep=0] (image) at (0,0) {\includegraphics[width=\textwidth, trim=8cm 6cm 9.08cm 14cm, clip]{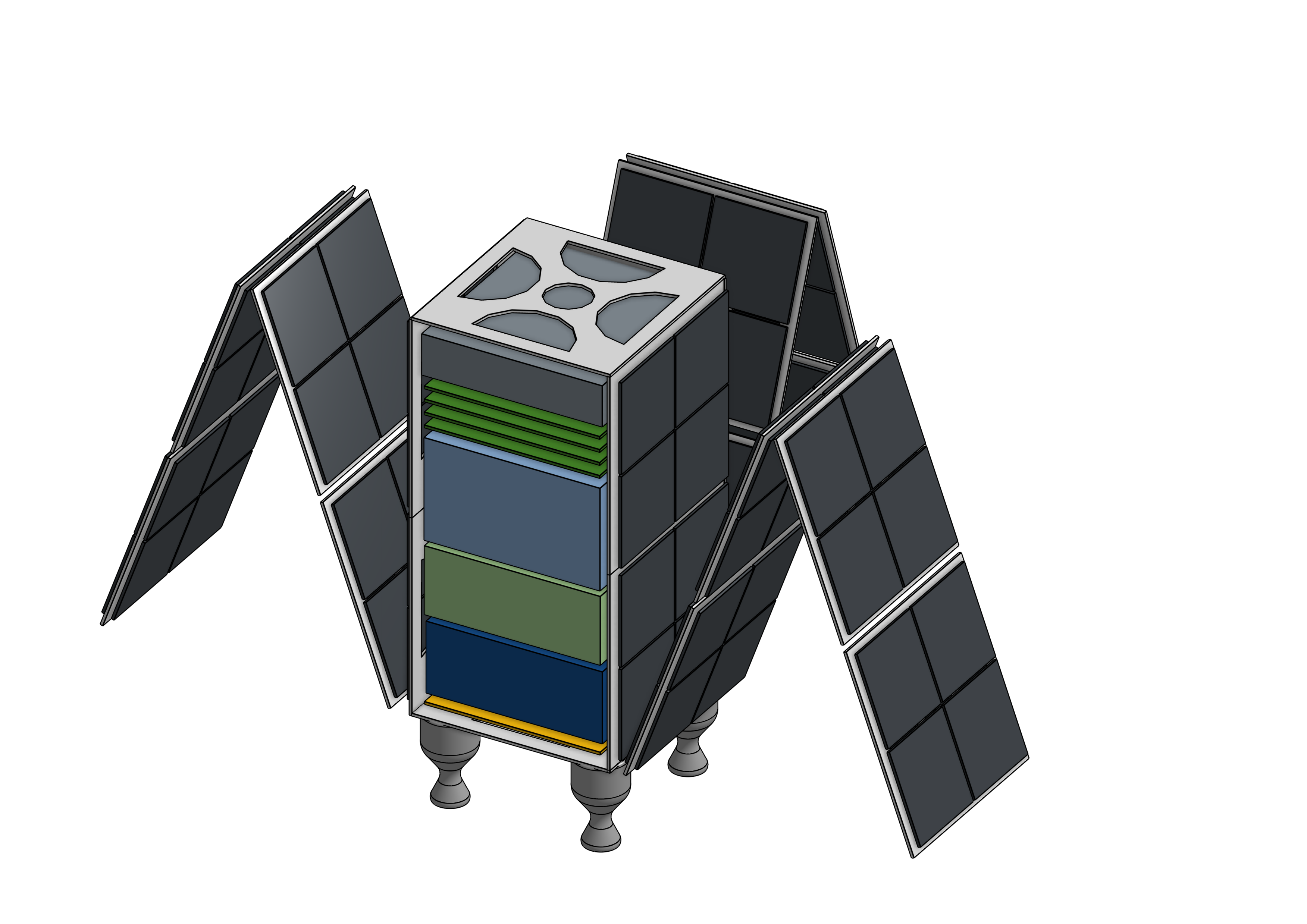}};
            \begin{scope}[x={(image.south east)},y={(image.north west)}]
                \draw[-{Latex[length=1mm,width=1mm]}] (0.4,1.0) -- (0.4,0.92) node[near start, above, yshift=1mm, align=left] {Stowed Antenna};
                \draw[-{Latex[length=1mm,width=1mm]}] (0.78,0.82) -- (0.72,0.74) node[near start, above, yshift=1mm, align=left] {Solar Panels};
                \draw[-{Latex[length=1mm,width=1mm]}] (0.01,0.78) -- (0.3,0.66) node[near start, above, yshift=2mm, xshift=-8mm, align=left] {Electronic Stack};
                \draw[-{Latex[length=1mm,width=1mm]}] (0.02,0.54) -- (0.3,0.54) node[near start, left, xshift=-6mm] {Battery};
                \draw[-{Latex[length=1mm,width=1mm]}] (0.18,0.3) -- (0.3,0.4) node[near start, left, yshift=-3mm, xshift=0mm, align=left] {CosmoCube Controller};
                \draw[-{Latex[length=1mm,width=1mm]}] (0.2,0.17) -- (0.3,0.26) node[near start, left, yshift=-1mm,xshift=-2mm, align=left] {Communication Box};
                \draw[-{Latex[length=1mm,width=1mm]}] (0.2,0.12) -- (0.32,0.2) node[near start, below, yshift=-1mm, xshift=-3mm, align=left] {Patch Antenna};
                \draw[-{Latex[length=1mm,width=1mm]}] (0.54,0.04) -- (0.49,0.04) node[near start, right, xshift=1.5mm, align=left] {Thrusters};
            \end{scope}
        \end{tikzpicture}
        \caption{CosmoCube in stowed configuration}
        \label{fig:cosmocube_annotated_b}
    \end{subfigure}
    \caption{CosmoCube configurations: (a) showcases the CosmoCube with its antenna deployed for the 21-cm signal detection, illustrating its readiness for operational mode in space. (b) presents the CosmoCube in its stowed configuration, highlighting the compact design with integrated solar panels and thrusters, indicating its transit and deployment efficiency. Note: The scale of individual instruments may vary.}
    \label{fig:cosmocube_combined}
\end{figure*}

To investigate the cosmic Dark Ages and capture the significantly redshifted 21-cm hydrogen line, a robust radiometer is required, capable of detecting faint, low-frequency signals ranging from around 10 MHz at a redshift of 150 to approximately 45 MHz at a redshift of 30. This device integrates wideband design, employs low-noise amplification, and implements precise calibration techniques \citep{Rogers2008}, with long integration times essential for accumulating sufficient data for a detectable signal-to-noise ratio. 

Building on this technological foundation, current experiments focus on redshifts from 6 to 28, targeting the EoR and CD. Our CosmoCube experiment aligns with the growing emphasis on lower frequencies, further extending the exploration of high redshifts, specifically targeting the range from 13 to 150 using a lunar orbiting CubeSat to detect the global 21-cm signal, as shown in Fig.~\ref{fig:CosmoCube coverage}. This approach not only promises insights into primordial fluctuations, non-Gaussian statistics deviations, and primordial gravitational waves but also reveals the cosmological model, dark matter properties, and the early Universe's conditions through the intergalactic medium's absorption against the CMB \citep{Pritchard2012}. Additionally, by being sensitive to electromagnetic energy injection, the Dark Ages' 21-cm signal offers a novel way to investigate dark matter models and explore exotic physics beyond current constraints, aiming for significant improvements in data precision \cite{Koopmans2021}.

The study by \cite{Mondal2023} provides a precise methodology for probing exotic dark matter models through the analysis of the 21-cm signal from the dark ages, focusing on variations such as millicharged dark matter, dark matter with non-gravitational interactions, and dark matter decay. These models have the potential to significantly alter the thermal and ionization history of the early Universe. By analyzing the power spectrum and global signal variations at high redshifts, models can be constrained, revealing how dark matter interacts with baryonic matter and impacts cosmic structure formation before the appearance of the first stars and galaxies. Although power spectrum analysis provides detailed insights into spatial variations and could reveal subtle dark matter influences, it requires complex observational setups. Due to simpler experimental needs, focusing on the global 21-cm signal is more practical initially, allowing measurements with a single antenna and paving the way for future, more detailed power spectrum studies.

The CosmoCube consists of various novel systems to generate a robust satellite described in the coming chapter. The experiments outlined in this paper extend to the characterization of a compact spectrometer utilizing the Xilinx ZCU111 evaluation board \citep{ZynqRFSoC2023}, which houses the XCZU28DR-2FFVG1517E Radio Frequency System-on-Chip (RFSoC). The selection of this RFSoC for the CosmoCube project's spectrometer represents a strategic decision, leveraging the device's integration of RF Data Converters, Analog-to-Digital Converters (ADCs), and Digital-to-Analog Converters (DACs) within RFSoC FPGA architecture for precise and efficient spectral analysis for isolating the 21-cm signal from the foreground noises. The compact RFSoC technology suits the cube satellite format, reducing size and power requirements, simplifying operational complexity, and lowering mission costs. By minimizing size and power requirements, the mission can utilize smaller, less expensive launch vehicles and extend operational life, thereby lowering overall mission costs. 

The RFSoC's commercial availability and its broad adoption within engineering and research communities offer access to shared knowledge and resources, enhancing development processes. However, reliance on commercially available technology and community resources introduces dependencies on external developments, potentially constraining project-specific innovations. Additionally, the absence of a radiation-hardened package for this product poses challenges for space applications, necessitating careful management of these limitations. However, \cite{Davis2019} successfully characterized the radiation hardness of the ZCU111, focusing on its resistance to single-event upsets (SEU) and single-event latch-up (SEL) in a proton irradiation environment, using protons with energies ranging from 60 MeV to 200 MeV. The findings confirm the device's potential for applications in space, particularly in low Earth orbit, highlighting its resilience to radiation-induced errors and the effectiveness of its design modifications in managing latch-up events, with latch-ups occurring at an impressively low rate of approximately once every 90 years. The findings confirm the device's potential for applications in space. However, further testing is advised to fully understand whether its radiation resistance also holds in the harsher radiation environment of lunar orbit, especially regarding its high-speed analog components.

Section 2 outlines the CosmoCube framework and lunar far side observations. Section 3 details the spectrometer characterization experiments. Section 4 evaluates FPGA capabilities for mission applicability. Section 5 concludes with final remarks.

\section{The far side of the Moon}

Deploying a radio telescope on the Moon's surface involves multiple challenges, such as uncertain lunar ground conditions and the ground spillover effect, which complicates signal detection. Using the Finite Difference Time Domain (FDTD) technique, \cite{Bassett2020} effectively characterized low-frequency radio wave propagation on the lunar far side. Building on this, \cite{burns2021} mapped the radio-quiet regions, enhancing the potential for precise observations from lunar missions and mitigating RFI from Earth. Despite these advancements, RFI from other lunar landers and orbiting satellites continues to pose significant challenges, as the effects of such interference are not fully understood due to the limited exploration of the Moon’s far side. The complexity of achieving a soft landing further complicates surface-based missions.

Given these challenges, space agencies' plans to return to the Moon make lunar missions increasingly advantageous and popular among researchers. Leveraging the interest and resources directed towards lunar exploration can facilitate the deployment of advanced scientific instruments.

Deploying an orbiting satellite offers considerable advantages. Satellites provide enhanced mobility and flexibility, allowing for dynamic observation strategies without the limitations of a fixed position. Additionally, they carry a lower risk of environmental interference compared to surface instruments, which are susceptible to dust and extreme thermal conditions that can impair functionality and longevity. However, an orbital approach also has disadvantages, such as increased susceptibility to RFI from the lunar surface, limited observation windows per orbit, potential systematics caused by reflections from the lunar surface interfering with the direct signal, and communication delays with the payload. Consequently, while an orbital approach simplifies operational demands and proves more cost-effective and efficient for conducting lunar observations, these challenges must be carefully managed.

For satellite-based telescope operations, establishing a direct line of sight with Earth facilitates communication, eliminating the dependency on intermediary relay satellites required by far side landers. This direct transmission method simplifies data exchange and significantly boosts the mission's overall efficiency.

To optimize the satellite's performance during nighttime operations, when the 21-cm signal detection is most favorable due to minimal solar noise and maximum Earth's occultation shielding from terrestrial RFI, careful planning of the orbital distance is critical, which set to be chosen in the range of 100 - 1000 km above the lunar surface. The integration of solar panels and precise sizing of the onboard battery ensure that the CosmoCube utilizes stored energy effectively. However, it is important to note that solar panels can also cause distortions in the EM field and beam of the instrument, necessitating careful design considerations. Additionally, having electronics in close proximity to the receiver requires extreme care in terms of shielding to prevent interference and maintain the integrity of the 21-cm signal detection. These operations necessitate a meticulous balance between energy harvest, storage, and usage, ensuring the satellite's functions are supported through the lunar night. This balance is achieved by carefully calibrating power generation and consumption to within a specific margin, ensuring that the satellite remains operational even under the most challenging thermal and power conditions. Thermal analysis for a 12U satellite, where '12U' refers to the satellite's size being equivalent to 12 standard CubeSat units (each 'U' typically being a 10x10x10 cm cube), was carried out under three scenarios: high solar flux during daylight, low solar flux during daylight, and a lunar eclipse, with reference data from a report by NASA (\citeyear{NASA2019DSNE}). The lowest temperature recorded during the lunar eclipse was -112°C. The extreme temperature variations on the Moon demand a sophisticated thermal management system. This system is crucial to protect the satellite’s electronics and battery, maintain component temperatures within operational ranges, and ensure robust calibration of the system, ensuring that the 21-cm signal is accurately estimated from the captured data.

One of the earliest missions was the Soviet Union’s Luna 3 in 1959. Luna 3 was the first spacecraft to photograph the Moon’s far side, returning grainy but groundbreaking images that revealed a landscape very different from the familiar near side \citep{Lipsky1962}. Another orbiter, Lunar Reconnaissance Orbiter (LRO), was developed by NASA in 2009, and the pictures are published online as open source \citep{Tooley2010}. China’s Chang’e-4 mission, launched in 2018, marked another significant milestone. Chang’e-4 was the first mission to land on the lunar far side, deploying a lander and a rover, Yutu-2, in the South Pole-Aitken Basin. This mission has provided detailed on-the-ground observations and measurements of the lunar far side, providing valuable data for lunar science and potential future manned missions \citep{Li2019}. 

More recently, India’s Chandrayaan-2 mission \citep{chand2}, launched in 2019, attempted a soft landing near the lunar south pole, but the Vikram lander unfortunately crashed during its descent. Building on this experience, India successfully landed its Chandrayaan-3 mission near the lunar south pole in 2023, marking a significant achievement in lunar exploration \citep{Chandrayaan3ISRO}. Japan has also been active in lunar exploration with its OMOTENASHI mission, launched in 2022, which aimed to be the first Japanese lunar lander. However, the mission failed due to communication issues, resulting in an unsuccessful landing \citep{OMOTENASHI19}. Russia’s Roscosmos also renewed its lunar exploration efforts with the launch of the Luna 25 mission in 2023, intended to land near the lunar south pole. However, the mission ended in a crash landing during descent, highlighting the challenges of lunar exploration \citep{Mitrofanov2021Luna25}. Another notable success was China’s Chang’e-5 mission in 2020, which not only landed on the Moon but also successfully returned lunar samples to Earth \citep{Zhou22}. Although several missions have been made to the far side of the Moon as yet, none have specifically aimed to detect the 21-cm signal.

The CosmoCube mission parameters, as specified in Table~\ref{table:Science mission requirements.}, aims to detect the 21-cm hydrogen signal, emphasizing the necessity for precise detection across 0 - 128 MHz frequency bandwidth, which covers the science goal frequency range of 10 MHz to 100 MHz. The channelizer's bin width helps reduce the noise floor of the system whilst increasing the power consumption. Thus, a compromise was made to find an optimum value of 62.5 kHz. This will be detailed in the later chapters. The clock frequencies are carefully synchronized and chosen to facilitate the identification of potential signal components from external sources within our targeted frequency range.

\begin{table}
\caption{The requirements for the science mission and radiometer are given in this table.}
\centering \footnotesize
\label{table:Science mission requirements.}
\begin{threeparttable}
\begin{tabular}{l c}
\toprule
Conditions & Goals and requirements\\ 
\midrule
Science goal for detection of 21-cm signal & 10-100 MHz \\
Total observed bandwidth & 0-128 MHz \\
Channel bandwidth & 62.5 kHz\tnote{*}  \\
The flatness throughout spectrum & $\pm$0.1 dB\tnote{*}\\ 
RF data converters’ matched clock frequency & 256 MHz \\
Minimum sensitivity of radiometer & 10 mK (30 < z < 250)\\
Mission 21-cm signal total observation time & minimum 1000 hours\\
Total mission duration & up to 3 years\tnote{*}\\
Orbit altitude & 100-1000 km\tnote{*}\\
Payload size & 12U or 16U\tnote{*}\\
\bottomrule
\end{tabular}
\begin{tablenotes}
\item[*] To be defined and justified by the science team.
\end{tablenotes}
\end{threeparttable}
\end{table}

As the response received at the antenna or the components in the front end does not have a flat response through our frequency of interest, a requirement of $\pm$0.1 dB is added. 

The 21-cm signal is generated using GLOBALEMU \citep{Bevins2021}. The constraining power of the CosmoCube data set depends on the shape and depth of the 21-cm signal relative to the noise in the data. Precision constraints are placed on parameters such as star formation efficiency (\(f_*\)), X-ray efficiency (\(f_x\)), minimum virial circular velocity (\(V_c\)), cosmic microwave background optical depth (\(\tau_{\text{CMB}}\)), and minimum X-ray frequency (\(\nu_{\text{min}}\)) in a high signal-to-noise regime, as demonstrated in REACH experiment \citep{Acedo2022}. This analysis is carried out using the Bayesian framework developed by \citep{Anstey2021}. The confidence in recovering these signals, measured by the difference in Bayesian evidence, increases significantly when the thermal sensitivity reaches 10 mK or better.

Achieving this sensitivity across the 5 - 45 MHz range with a 0.1 MHz channel width requires approximately 1000 hours of integration, as evaluated by the radiometer equation. This integration time is the minimum needed to accurately reduce thermal noise for reliable estimation of the 21-cm signal. With a mission duration of up to three years, there is flexibility for further optimization to ensure the success of the observations.

The 10 mK limit restricts the ability to constrain certain physical parameters. \cite{Mondal2024} demonstrate the relative errors and limits on various parameters across different noise temperature levels, which increase with higher redshifts. The baryon density ($\Omega_b$) and total matter density ($\Omega_m$) can be reliably estimated in the bandwidth. In contrast, there is greater uncertainty in constraining the helium fraction ($Y_P$). Our aimed thermal sensitivity aligns with the difference between the two experiments, where the signal strength predicted by Planck’s measurement of the reduced Hubble constant ($h$) \citep{Planck2018} is about 10 mK lower than that inferred from the SH0ES measurement \citep{Riess_2022} within the $\Lambda\text{CDM}$ model.

Key considerations for the orbital altitude include the use of Earth occultation to minimize RFI, establishing reliable communication between the payload and Earth, and addressing the cost of satellite end-of-life disposal, which increases with orbit altitude due to the higher fuel or energy required for controlled re-entry. According to \cite{itu_radio_astronomy_2013}, man-made RFI sources radiate from up to 100,000 km from Earth's center. This creates a cone-shaped shielded zone on the far side of the Moon, extending about 6,700 km from the lunar center, with the Moon’s diameter being approximately 3,500 km. Earth-based RFI observations by \cite{Yan2023} reported a maximum 2 dB increase in their measured spectrum of -147 $\text{dBm}/\text{Hz}$, which includes sky and receiver signal between 6 and 20 MHz in the shadow of the Moon's far side. This suppression of RFI aligns with the dynamic range requirement of around 80 dB to effectively study cosmological signals from the Dark Ages.

The highest sky brightness temperature within the bandwidth is around $10^{5}$ K, approximately 70 dB above our targeted thermal sensitivity. While an 80 dB dynamic range is sufficient to enable the analysis of the Dark Ages signal from the brightest sky temperature, stronger RFI around the Moon, which is not included in this estimate, may require a greater dynamic range. The 80 dB estimate provides adequate headroom between the strongest sky signal and the targeted weakest measurement, 10 mK. Although achieving this dynamic range is feasible, achieving the 10 mK sensitivity will require rigorous calibration, a process currently under study. Future work will focus on calibrating the system with sufficient stability to maintain 10 mK sensitivity over extended periods.

The initial model of CosmoCube is shown in Fig.~\ref{fig:cosmocube_annotated_a} for the deployed satellite structure and in Fig.~\ref{fig:cosmocube_annotated_b} for the stowed satellite structure. To meet its operational energy requirements, the CosmoCube is equipped with solar panels and a battery pack designed to provide a continuous power supply for the duration of the mission. This setup ensures all satellite components receive regulated power for sustained optimal performance. The satellite's orientation and stability in lunar orbit are meticulously controlled by altitude and thruster management systems, facilitating precise maneuverability and required elements to deorbit at the end of its lifecycle.

Central to the CosmoCube's scientific payload is a radiometer, supported by a deployable science antenna for the 21-cm signal acquisition, an RF front end for initial signal processing, and a calibrator to enhance measurement accuracy. The stowed antenna is designed to fit in about 1U package during the launch. One of the elects of the deployed antenna is about 3 m long. Besides the bowtie, the other types of candidate antennas include dipole and blade dipole. However, this paper will not delve into the details of the antenna design, as it is still under development. Regardless, it is important to note that the presence of the satellite body and solar panels can significantly affect the beam shape by introducing distortions and reflections that alter the intended radiation pattern. These distortions can lead to inaccuracies in the 21-cm signal acquisition, potentially complicating data interpretation. Careful consideration and mitigation strategies, such as type of antenna, precise placement of the antenna relative to the satellite structure, and advanced EM modeling, will be crucial to minimize these effects.

\section{System Architecture and Instrumentation}

\begin{figure}
  \centering
    \includegraphics[width=\columnwidth, trim=2cm 7.4cm 1.6cm 11.5cm, clip]{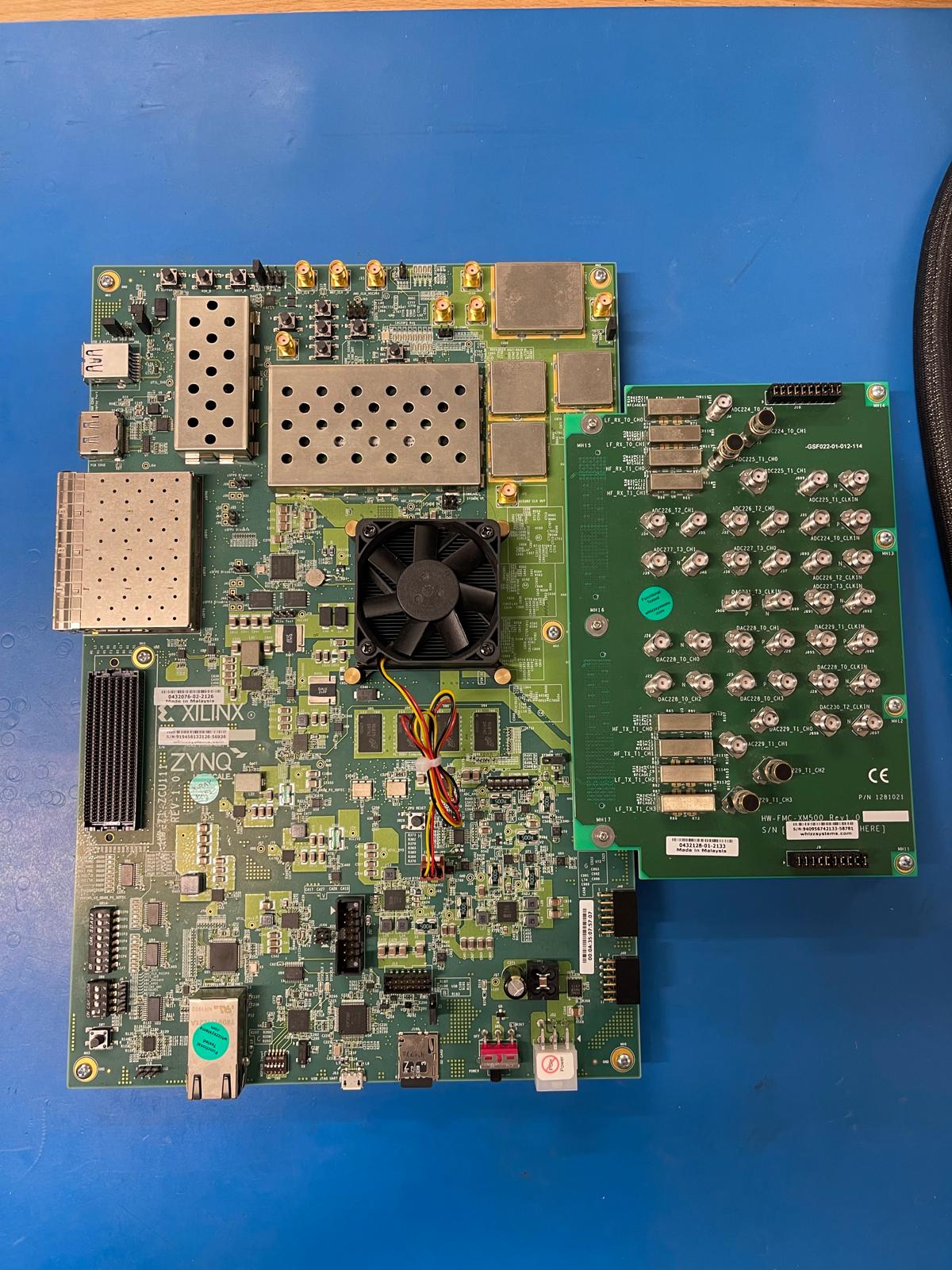}
    \caption{The ZCU111 Evaluation Board is configured for 21-cm signal detection experiments, interfacing with an XM500 balun board equipped with SMA connectors.}
    \label{fig:ZCU111}               
\end{figure}

This paper delves into the capability of the RFSoC FPGA, pictured in Fig.~\ref{fig:ZCU111}, for efficient and precise spectral analysis, which is essential for isolating the 21-cm signal from foreground noise. The compact design of the RFSoC is particularly beneficial for the cube satellite format of the CosmoCube project. This integration not only reduces the physical size and power requirements of the satellite but also cuts down on operational complexity and overall mission costs.

\begin{figure}
    \centering
    \includegraphics[width=\columnwidth,trim=0cm 0cm 0cm 0cm, clip]{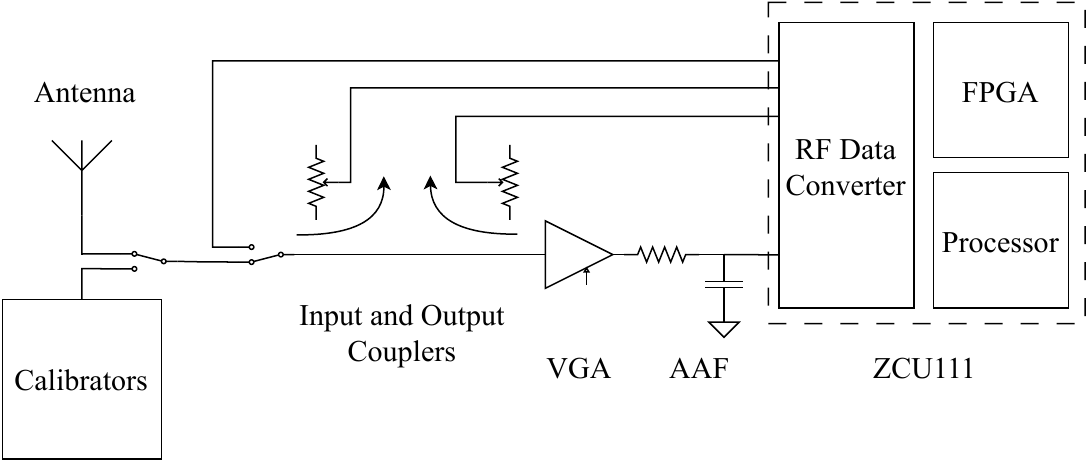}
    \caption{This diagram depicts the radiometer with signal flow from the antenna through to the ZCU111 board, with various front-end components, including an amplifier, anti-aliasing filter, and couplers. It illustrates the incorporation of three ADCs and one DAC within the RFSoC, which are pivotal for the precise translation and processing of the 21-cm signal for astrophysical analysis and calibration of the radiometer.}
    \label{fig:front-end}
\end{figure}

Fig.~\ref{fig:front-end} illustrates CosmoCube's front-end setup, featuring calibrators parallel to the antenna to simulate observations and generate various noise profiles. This setup, using a Dicke switch-based calibrator system similar to the one in a paper by \cite{Nima2023}, adopts Vector Network Analyzer (VNA) principles for calibration. This includes input and output couplers with a switching mechanism for seamless calibration and detection mode transitions. A Variable Gain Amplifier (VGA) compensates for nonlinear gain responses from the antenna and other components, enhancing signal linearity within the measurement bandwidth. An anti-aliasing filter (AAF) is incorporated to meet Nyquist's theorem, ensuring accurate signal sampling by avoiding aliasing.

For this setup, three ADCs and one DAC are utilized within the RFSoC. DAC generates waveforms for calibration purposes, whilst two ADCs measure power on the incident and reflected signals on the couplers to measure the reflection coefficients. The final ADC generates a spectrum by processing ADC samples through a channelizer. We have utilized Simulink blocks to design FPGA algorithms.

\begin{figure*}
    \centering
    \includegraphics[width=\textwidth, trim=0cm 0cm 0cm 0cm, clip]{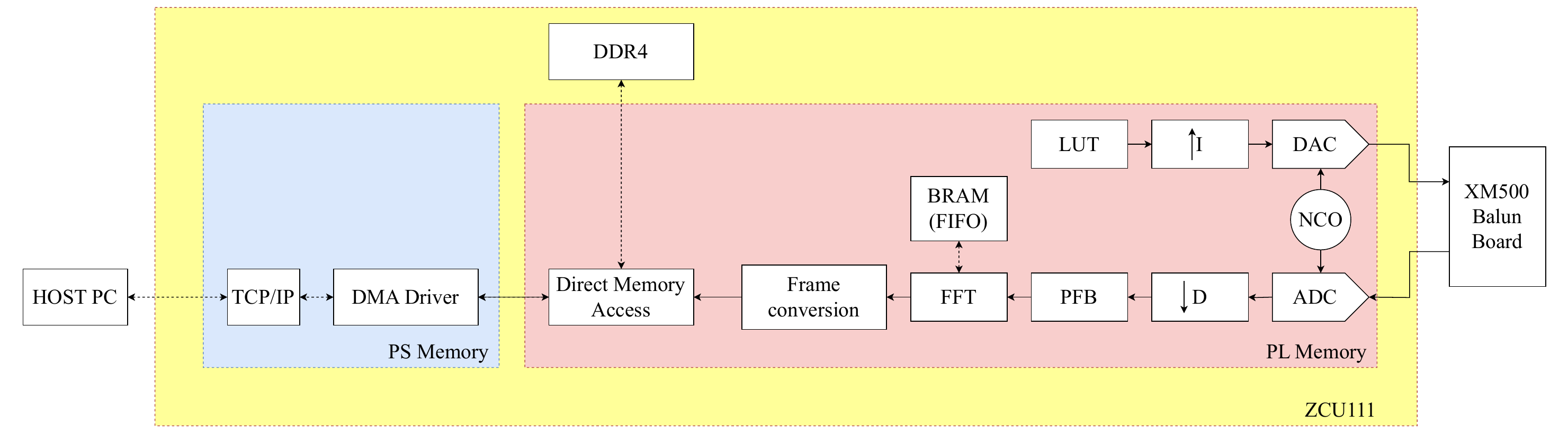}
    \caption{This architecture diagram showcases the data flow within the ZCU111 RFSoC, starting from the ADC and DAC stages through to the processing and memory elements. AXI Stream is utilized to convey digitized samples within the programmable logic (PL). AXI Lite is used to communicate with the AXI interface to interact with the processing system (PS).}
    \label{fig:rfsoc_diagram}
\end{figure*}

In the detection mode, the system operates with a single ADC active, while the calibration mode activates two extra ADCs and a DAC. The configuration of single ADC and single DAC is detailed in Fig.~\ref{fig:rfsoc_diagram}. In the receiving operation, the signal's journey through the architecture starts with its capture by the XM500 Balun Board, followed by digitizing the signal via the ADC. Post-ADC, the signal undergoes decimation to reduce the data rate for more manageable processing while maintaining the integrity of the signal information. The Polyphase Filter Bank (PFB) then efficiently separates the signal into multiple frequency bands, optimizing the signal for the Fast Fourier Transform (FFT) process. The FFT block transforms the time-domain data into the frequency domain, facilitating the identification and analysis of specific frequency components. Following this transformation, data is temporarily stored in Block RAM (BRAM), configured as First In First Out (FIFO) for efficient data management. Following FFT, data is routed through buffers and Direct Memory Access (DMA) for efficient data handling, eventually being stored in DDR4 memory. The processed data is then accessible for further analysis externally, completing the reception and initial processing phase, ensuring a comprehensive approach to cosmic signal detection and analysis.

The process of PFB involves segmenting a high-rate data stream into multiple lower-rate streams using input commutators, enhancing manageability and reducing complexity. Each channel is processed by an FIR (Finite Impulse Response) filter with varying coefficients to optimize signal integrity and minimize distortion. This setup manages spectral folding or aliasing when reducing sample rates, ensuring precise frequency band extraction. A subsequent FFT phase aligns and separates the baseband aliases, shifting signals from the time domain to the frequency domain for detailed analysis or further processing. This integrated approach significantly enhances both efficiency and performance in digital communication systems.

Using a Look-Up Table (LUT) for predefined signal patterns, the DAC precisely converts digital signals into analog form. These signals are then transmitted through the XM500 Balun Board to aid in system calibration. The waveform generation starts with the Numerically Controlled Oscillator (NCO), which produces a precise frequency. This frequency is interpolated to enhance the Signal-to-Noise Ratio (SNR) before being modulated and fed back into the DAC for final analog conversion.

Given the importance of precise signal processing in our system, we carefully considered the trade-offs between floating-point and fixed-point arithmetic. Floating-point operations provide a wide range of values and precision, but they require more resources and have higher latency. Furthermore, floating-point numbers are prone to precision loss due to their limited representation of real numbers. This limitation becomes pronounced when dealing with large values, leading to information loss in significant digits. Conversely, fixed-point operations are more resource-efficient and have lower latency, although they have a limited range and precision. While it offers less dynamic range compared to floating-point arithmetic, because of the rounding errors inherent in fixed-point calculations. However, for our channelizer, where efficient resource utilization is crucial, fixed-point arithmetic is utilized due to its lower overhead. For the post-processing of the data, we worked with floating-point arithmetic on Matlab to plot the measurements. 

An interpolated signal, when presented at a higher sampling rate, is an exact reproduction of the original signal. Upsampling a signal sampled at F to a higher sample rate (N*F) involves inserting N-1 zeros between each sample of the original signal. This process results in N replications of the original spectrum due to the creation of spectral images at multiples of the original sampling frequency. Low-pass filtering of the result helps to remove unwanted replications resulting in the preservation of the original signal at the required higher sample rate.

Various testing equipment was utilized to characterize the RFSoC for low frequency spectrometer design. The Keysight E8257D generated signals to assess the harmonic distortions and noise performance of the RFSoC ADC. The Keysight PNA-X was used to analyze the data generated by the RFSoC DAC, specifically measuring the Spurious Free Dynamic Range (SFDR) in spectrum analyzer mode. Furthermore, the Keysight PNA-X was used to generate two-tone signals.

In the CosmoCube project, the ZCU111 RFSoC FPGA's DAC setup with a sample rate of 2048 MSPS for 65536 samples per clock cycle. A 256 MHz clock frequency was chosen to cover the necessary band up to 128 MHz. Although our mission targets frequencies up to 100 MHz, the 128 MHz bandwidth provides a margin for signal processing and filtering, addressing potential out-of-band noise or interference. A higher ADC clock frequency helps avoid aliasing and accurately sample the signal, ensuring stable observation and acquirement of the spectrum with the channelizer. The 32-bit stream data width provides a significant dynamic range, ensuring the preservation of signal integrity. Since this exceeds the RFSoC ADC's bit depth of 12-bit, it maintains data integrity and allows for additional data to be streamed concurrently.

The samples are transmitted to an external source at a data rate of 655 kbps. This is a relatively high speed, requiring substantial data storage in the payload before transmission to Earth. Assuming the payload is at 1000 km above the Moon, we have 43 minutes of data collection during Earth's occultation, which is the area inside the cone-like shadow of the Moon. At a data rate of 655 kbps, this results in approximately 201 MB of data, which is significant. However, due to averaging within FPGA, this data will be reduced to below 4 MB per observation on the far side.

\section{Hardware Characterization}

This section outlines the methodology for measuring the accuracy, efficiency, and overall performance of the RF data converters through a series of systematic tests. This analysis outlines testing methods for assessing the performance of these converters in capturing low-frequency signals from the dark ages. Accurate characterization ensures that the RFSoC meets the sensitivity and dynamic range required for studying cosmic phenomena, with performance parameters optimized to reflect the variations in signal properties across different redshift.

\subsection{Single tone measurements}
\label{sec:singleTone}

In an initial characterization, the DAC and ADC of the RF Data Converter of the RFSoC are assessed by single-tone measurements within the bandwidth of 0 to 128 MHz. However, the characterization of this low frequency introduces challenges in distinguishing spurious signals generated by the device under test, ZCU111, and the testing equipment used in this experiment. The commercially available spectrum analyzers trigger harmonic generation attributed to the first up/down converter mixers in its signal chain saturating. A substantial fraction of our observed spectrum consists of harmonic frequencies of fundamental signals. Consequently, it is essential to carefully control mixer input levels to prevent overdriving. We employed various notch filters to differentiate harmonics produced by the DACs of RFSoC from the spectrum analyzer PNA-X. Additionally, an anti-aliasing filter with a cutoff frequency of 105 MHz was utilized to remove aliased frequency components that could otherwise distort the analysis within our bandwidth. In contrast, various combinations of low-pass and high-pass filters built in-house are utilized in assessing RFSoC's ADC performance as they eliminate harmonics generated by the signal generators. The list of used filters is given in Table~\ref{table:RFfilters}.

\begin{table}
\caption{Below presents the range of RF filters, with their respective cutoff frequencies and models, utilized in testing the RFSoC's ADC and DAC functionalities, highlighting both commercially available and in-house designed models.}
\centering \footnotesize
\label{table:RFfilters}
\begin{threeparttable}
\begin{tabular}{l c c}
\toprule
Cutoff frequencies & Filter model & Test used for\\ 
\midrule
9-15 MHz BPF & ZX75-12-S+ & ADC and DAC \\
22 MHz 9th Chebyshev LPF & designed in-house & ADC \\
26 MHz 9th Chebyshev LPF & designed in-house & ADC \\
37 MHz 9th Chebyshev LPF & designed in-house & ADC \\
62 MHz 9th Chebyshev LPF & designed in-house & ADC \\
105 MHz 9th Chebyshev LPF & designed in-house & ADC \\
25 MHz HPF & CHPFL-0025-BNC & ADC and DAC \\
40-250 MHz BPF & designed in-house & ADC and DAC \\
90 MHz HPF & BHP-100 & ADC and DAC \\
88-108 MHz BSF & BSF-108+ & ADC and DAC \\
160 MHz HPF & SHP-175+ & ADC and DAC \\
\bottomrule
\end{tabular}
\end{threeparttable}
\end{table}

\begin{figure}
    \centering
    \includegraphics[width=\columnwidth, trim=1cm 6cm 0.5cm 7cm, clip]{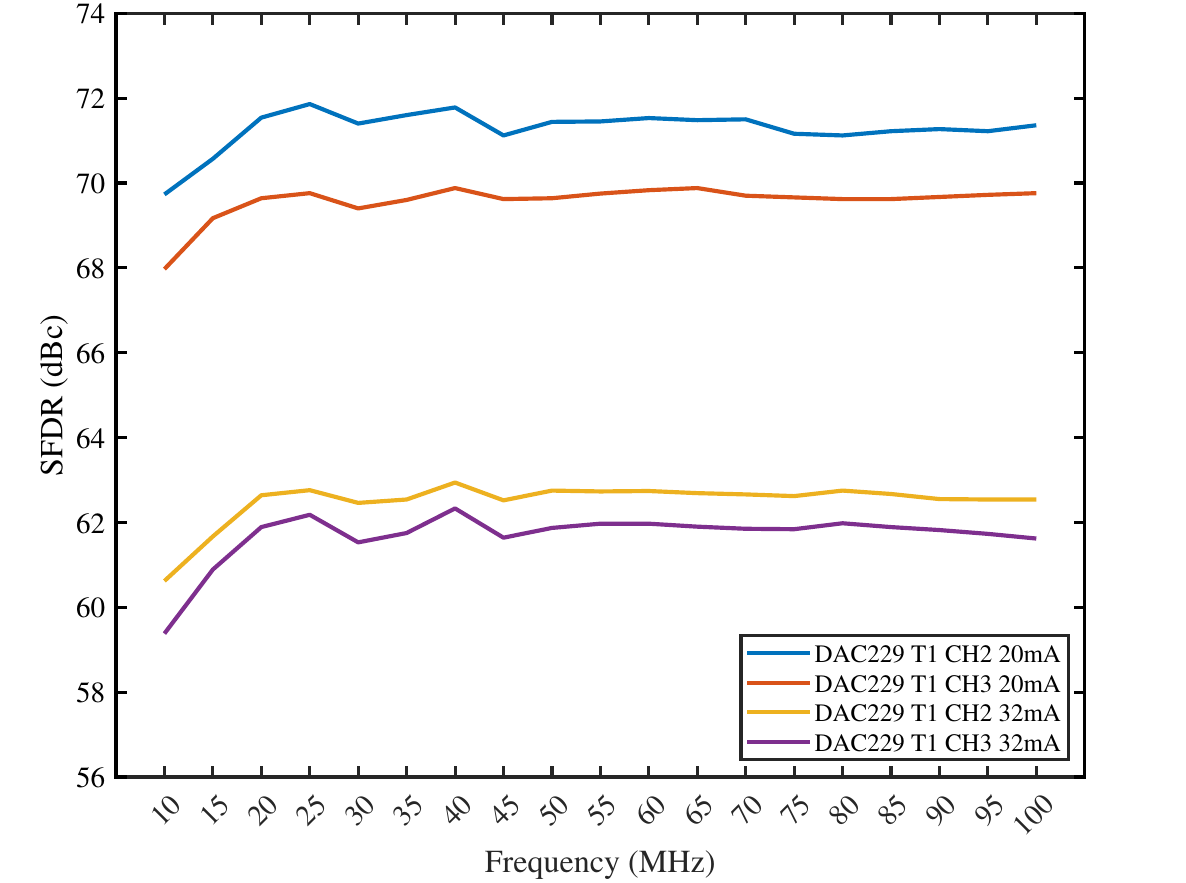}
    \caption{The SFDR of RFSoC's DAC-generated signals across varying frequencies, as measured by the Keysight PNA-X Network Analyzer in spectrum analyzer mode. The graph illustrates performance at different current settings, showing the SFDR peak at -3 dBm for the 20 mA mode and at +1 dBm for the 32 mA operation. This representation underscores the impact of operational current on SFDR, with an SNR-optimized selection for DAC.}
    \label{fig:SFDRrfsocDAC}
\end{figure}

The 14-bit DAC within the RFSoC can be optimized for low SNR or high linearity. SNR optimization focuses on maximizing the signal's power relative to background noise, employing noise reduction techniques to enhance signal clarity. Conversely, linearity optimization aims to maintain signal fidelity across its operating range, using strategies to ensure the output accurately reflects the input without distortion. Selection of SNR optimized decoder triggers a requirement of characterizing spurious signals within a bandwidth of 300 MHz for fundamental signals up to 100 MHz. The sample rate was chosen to be 2048 MSPS, and the clock frequency was set to 256 MHz. Fig.~\ref{fig:SFDRrfsocDAC} illustrates the SFDR characterization, the ratio between the fundamental and the next highest spurious signal excluding the DC component, of two DACs with two different decoder modes, low power with 20 mA and high power with 32 mA operations. The output power of -3 dBm is observed within the bandwidth with varying frequency for low power mode, whilst +1 dBm was acquired throughout the band for high power mode, holding for the same digital code in each case. Only about a 0.06 dBm decrease was observed in the fundamental signal powers throughout the frequency range. As the band-pass filter (BPF), high-pass filter (HPF) or band-stop filter (BSF) is used, the output power is matched to separate measurements of maximum values for the frequency to eliminate power losses introduced by the filters. The datasheet of RFSoC indicates another power mode, 3 V operation, requiring reconfiguration of the Infenion IRPS5401 power supply within the evaluation board with their programming tools providing +5 dBm output with a higher voltage setting. Hence, this was not investigated. Another limitation of the Simulink RF Data Converter block is that it only supports 20 mA operation, which was overcome by adjustment of this value in the Xilinx Vivado tool.

Although the RFSoC is integrated with eight 14-bit DACs, only two have been investigated as XM500 balun board consists of 2 to 1 Mini-Circuits TCM2-33WX+ transformers operating between 10 and 3000 MHz. While the other two DACs work with high-frequency applications, the last four are two port differential outputs. The two DACs, Tile229CH2 and Tile229CH3, show differences in their performances, which might be explained by external components such as RF transformers or SMA connectors, which are screw-on and can easily be disturbed as the connection changes each time in the daughter board. Although the SMA connectors were replaced, the performance differences persisted, likely due to the inherent fiddling and connection variability that remains even with new connectors. Another explanation is that the RFSoC chip shows these varieties within its structure. 

The SFDR was mainly dominated by the 2nd and 3rd harmonics generated by the DACs. The DACTile229CH2 demonstrated better performance, with about 71 dBc for low power and 63 dBc for high power mode. The datasheet of Xilinx shows SFDR measurements in a minimum of 240 MHz range and with a sampling rate of 6.4 GSPS, and compared to this, our values align with the datasheet values of 71 dBc for low power and 64 dBc for high power. Although the SFDR difference is apparent between the two DACs, the generated fundamental signal power was the same for both DACs. This shows that in order to implement robust calibration methods, we have to compensate for the difference between these ports.

\begin{figure}
    \centering
    \includegraphics[width=\columnwidth, trim=1cm 6cm 0.5cm 7cm, clip]{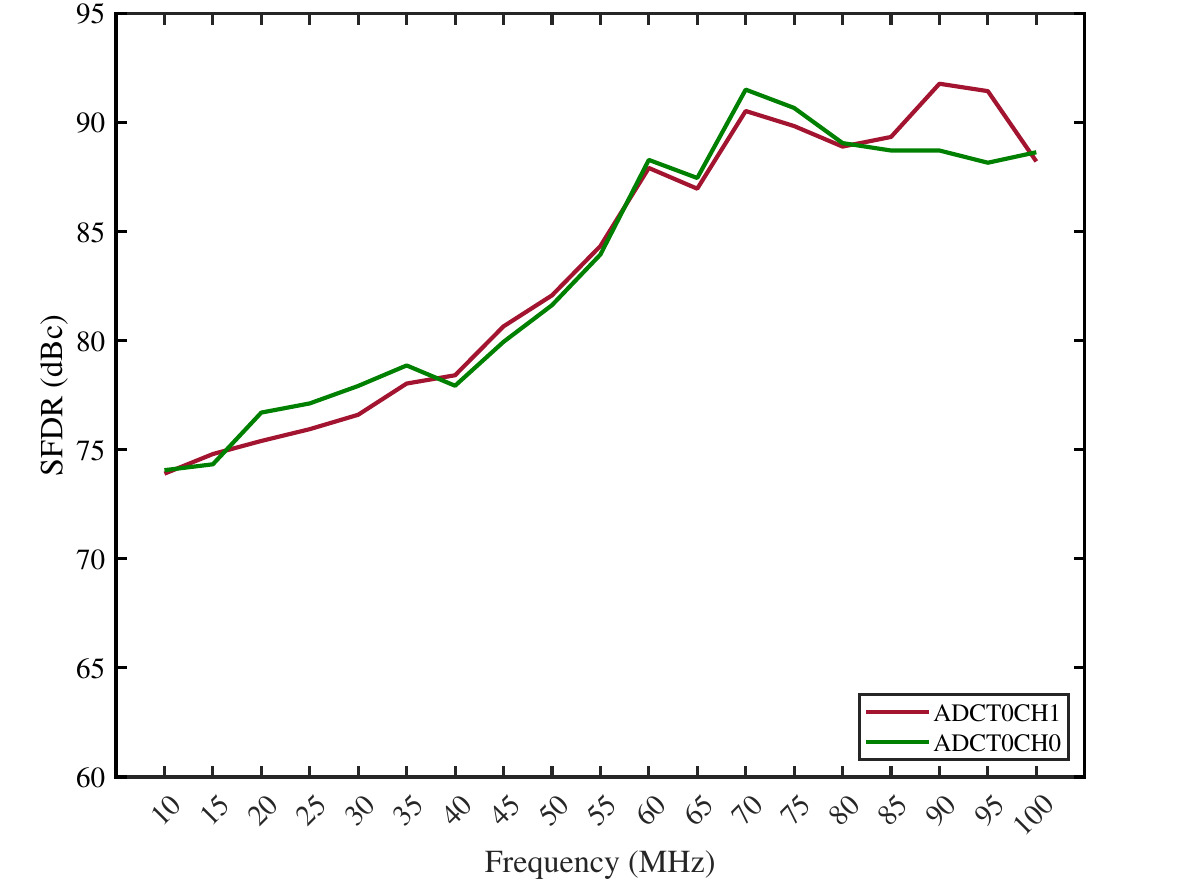}
    \caption{SFDR performance of ADCTile224CH0 and ADCTile224CH1 over frequency range. This graph displays the SFDR of two ADC channels obtained through a 65536-point FFT. The signals for this analysis were produced using a Keysight E8257D signal generator at -1dBFS, and various LPFs were applied to distinguish harmonics.}
    \label{fig:your_label}
\end{figure}

Various low-pass filters (LPF) were used in the measurement of the signal through ADC, as listed in Table~\ref{table:RFfilters}. Datasheet of \cite{ZynqRFSoC2023} indicates +1 Vpp as the full-scale range for the ADCs. The power levels for each measurement matched to -1 dBFS, corresponding to 3 dBm power for a 50 Ohms system, in order to eliminate clipping and compensate for the losses introduced by the filters. ADCTile224CH1 and ADCTile224CH0, separately, employ a 65536-point FFT to analyze incoming signals, with a channel bandwidth of 3.9 kHz with PFB. The clock signal of 256 MHz is employed with a 2048 MSPS sampling rate. 100-spectrum data was acquired, and the average was taken.

The improvement with the SFDR was observed with the higher frequencies, as the harmonics generated in the ADC fall out of our bandwidth from 65 MHz and above, creating their folded frequency components triggered in the Nyquist range. Similar to the RF transformers used in the balun board for DAC outputs, the same components are located in the ADC receive chain. Xilinx datasheet illustrates 2nd and 3rd harmonic distortions from -68 to -78 dBc at 240 MHz signal with -1 dBFS input and 2GHz clock frequency. The SFDR is similar to the one in the datasheet for low frequency, where harmonics are within our spectrum. Conversely, the SFDR excluding harmonics on the Xilinx datasheet indicates a typical 85 dBc at 240 MHz operation. We acquired about 90 dBc SFDR for the higher frequencies. The two ports' similar behavior and the slight differences might be explained as the on-chip variation of these legs or the traces leading up to them.

\subsection{Spectrometer design}

Following the characterization of DACs and ADCs within the RFSoC, we have developed a spectrometer with PFB to assess its qualification for CosmoCube. For adequate and power-efficient capture of the 21-cm signal from the Dark Ages, the PFB configuration utilized in this setup is designed to generate 81920 coefficients for each frame comprising 4096 samples, yielding a bin width of 62.5 kHz. The large number of coefficients allows for precise filtering and enhances the ability to isolate the faint 21-cm signal from stronger background noise and interference. The samples per frame provide a manageable data size that supports efficient processing while still achieving a bin width, which is fine enough to resolve the spectral features of the 21-cm signal within the Dark Ages. This configuration ensures that the signal's subtle variations can be detected and analyzed without excessive power consumption, which is crucial for long-duration observations in space-based missions.

\begin{figure}
    \centering
    \includegraphics[width=\columnwidth, trim=1cm 6cm 0.5cm 7cm, clip]{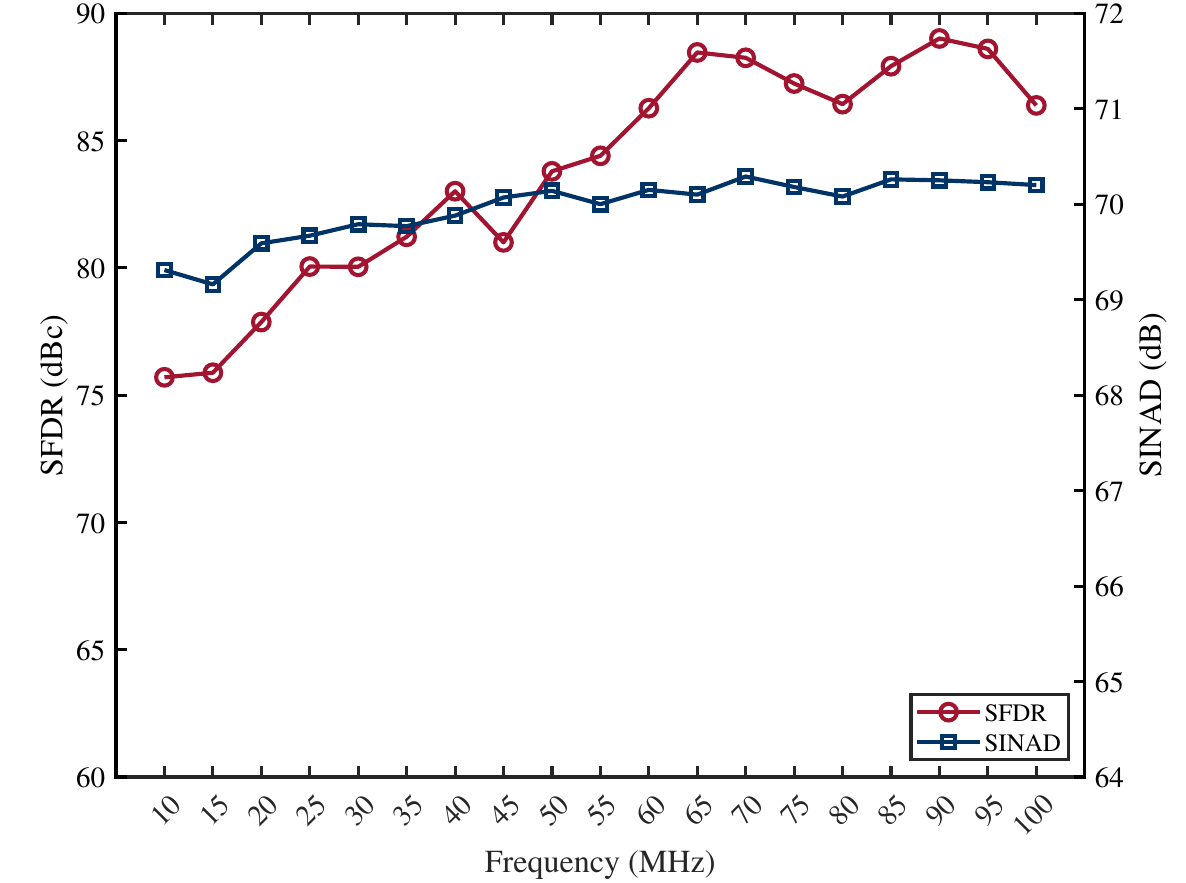}
    \caption{Illustrates the performance of the spectrometer through a 4096-point FFT, highlighting the SFDR and SINAD across frequencies ranging from 10 MHz to 100 MHz. Measurements were conducted using a signal generator and LPFs, capturing the variations of each metric for RFSoC separately from the test equipment.}
    \label{fig:specSINADandSFDR}
\end{figure}

The SFDR and Signal-to-Noise and Distortion Ratio (SINAD) of the designed spectrometer are illustrated in Fig.~\ref{fig:specSINADandSFDR}. SINAD encompasses both, noise and distortion, including spurious signals, but excludes DC relative to the fundamental signal's magnitude. The RMS of the fundamental signal reflects the signal's effective amplitude. The root-sum-square (RSS) of noise and spurious signal components provides a single figure for their cumulative effect on its quality. The continuous wave (CW) signals were generated by Keysight E8257D. The LPFs and HPFs were combined to create the BPFs. Although the BPFs mitigate the noise floor generated by the signal generator to present at the input of the ADC, it is very hard to filter the noise floor around the peak power. The fundamental power level measured at the ADC captured spectrum when input was -1 dBFS only changed by -0.22 dBFS at the highest frequency because of the quantization noise. Hence, this loss was ignored, and the fundamental power levels received at the ADC were matched over the frequency band.

The low-frequency SFDR was found to be similar to the previous results, with 65536 FFT length for the same ADC setting, and the levels of the harmonics were the same. And, these values are still compatible for a spectrometer to be used in a space-based radiometer. The SINAD remains relatively stable across the frequency range, indicating consistent overall signal integrity within our bandwidth and performance required in cosmic observations.

Table~\ref{tab:PowerConsumption} demonstrates the difference between the maximum available FFT length, as presented in the previous Section~\ref{sec:singleTone}, and the designed spectrometer presented here. The compromise is given in terms of power budget and robust signal detection. This table highlights differences in specifications set in the Matlab model, such as sampling rate, clock frequency, FFT length, and the number of coefficients in the PFB, as well as FFT bin width. Additionally, on-chip utilization metrics received through the Vivado tool, including Look-Up Tables (LUT), Block RAM (BRAM), Digital Signal Processing (DSP) utilization, and power consumption, are detailed, reflecting the efficiency and resource demands of each setup. Although a larger FFT bin width results in reduced spectral resolution, it enhances power efficiency, and the spectral resolution provided by the spectrometer is adequately sufficient for analyzing the global 21-cm signal. However, it is important to consider that low spectral resolution could pose challenges in the presence of RFI, which is often very narrow-band. Ground-based experiments have shown that insufficient spectral resolution can lead to significant data loss due to RFI contamination. Thus, while the designed spectrometer balances power efficiency and resolution, situations with strong RFI might justify the slightly higher power consumption associated with achieving higher spectral resolution to preserve data integrity.

\begin{table}
    \centering
    \caption{Two channelizer setups with maximum performance and spectrometer performance. In both cases, only the receive chain is active, except for the power consumption, which includes the operation of three ADCs and one DAC.}
    \label{tab:PowerConsumption}
    \begin{tabular}{lccr} 
        \hline
        \textbf{Specifications} & \textbf{Maximum} & \textbf{Spectrometer} \\
        \hline
        Sampling Rate & 2048 MSPS & 2048 MSPS\\
        Clock Frequency & 256 MHz & 256 MHz\\
        FFT Length & 65536 & 4096\\
        PFB Number of Coefficients & 262144 & 81920\\
        FFT bin BW & 3.9 kHz & 62.5 kHz\\
        \hline
        \multicolumn{1}{l}{\textbf{On-chip Utilization}} \\
        \hline
        LUT & 6.37\% & 3.27\%\\
        LUTRAM & 0.87\% & 0.49\%\\
        BRAM & 87.27\% & 23.75\%\\
        DSP & 1.57\% & 1.29\%\\
        Power Consumption & 6.1 W & 5.45 W\\
        \hline
    \end{tabular}
\end{table}

\subsubsection{Key metrics for ADC performance for spectrometer}

The spectrometer is characterized by typical metrics like Effective Number of Bits (ENOB) and Noise Spectral Density (NSD), which quantifies the converters' precision and noise levels, thereby influencing the overall fidelity of signal detection and analysis. Eq.~\ref{eq:enob} gives the ENOB, where the first component is the ratio of the fundamental signal to all the components of the noise and distortion, excluding DC. The second part, -1.76 dB, indicates the full-scale sine wave's quantization error in the 12-bit ADC. The final component of the equation, as given in \cite{AnalogMT003}, normalizes to the reduced power used compared to the full scale given in the datasheet. The denominator converts the binary resolution of the ADC into the dB scale.

\begin{equation}
    ENOB = \frac{10 \log\left(\frac{P_{\text{fund}}}{P_{\text{spurs}} + P_{\text{noise}}}\right) - 1.76 + 20 \log\left(\frac{FS_{\text{pp}}}{A_{\text{pp}}}\right)}{6.02}
    \label{eq:enob}
\end{equation}

where \( P_{\text{fund}} \) represents the power of the fundamental signal. \( P_{\text{spurs}} \) refers to the power of spurious signals, and \( P_{\text{noise}} \) is the power of all noise sources that degrade the signal. \( FS_{\text{pp}} \) is the full-scale peak-to-peak voltage, indicating the maximum input range of the ADC, while \( A_{\text{pp}} \) is the peak-to-peak amplitude of the actual input signal, representing its dynamic range within the converter's limits.

A wider measurement bandwidth will include more noise power, as the total noise is the integration of NSD across the bandwidth. Therefore, to maintain a high Signal-to-Noise Ratio (SNR) in the system design, we considered the bandwidth over which the spectrometer operates. NSD is calculated with Eq.~\ref{eq:nsd}, where the second component illustrates the limitation with a set sampling frequency, as shown in \cite{XilinxRF2017}. Since measurements compared with the signal are less than ADC's full scale utilized, the normalization factor appears in the equation.

\begin{equation}
    NSD = -10 \log\left(\frac{P_{\text{fund}}}{P_{\text{noise}}}\right) - 10\log\left(\frac{f_{\text{s}}}{2}\right) - 20\log\left(\frac{FS_{\text{pp}}}{A_{\text{pp}}}\right)
    \label{eq:nsd}
\end{equation}

where \( f_{\text{s}} \) is the sampling frequency of the ADC.

\begin{figure}
    \centering
    \includegraphics[width=\columnwidth, trim=0.2cm 6cm 0.5cm 7cm, clip]{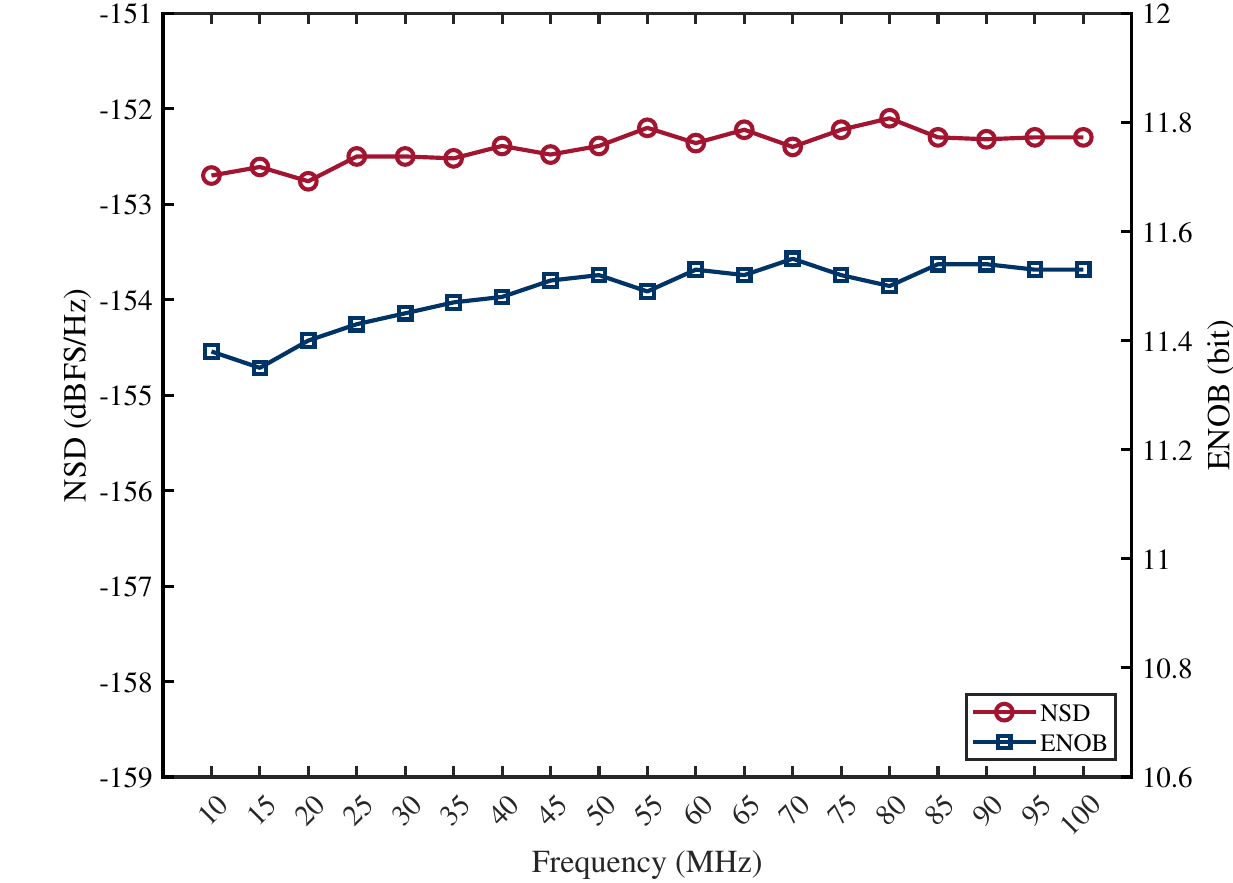}
    \caption{NSD and ENOB across frequency range for the spectrometer. This graph illustrates the performance of a 4096-point FFT analysis showing stable NSD and ENOB metrics as functions of frequency from 10 MHz to 100 MHz. The data here is the same measurements taken for Fig.~\ref{fig:specSINADandSFDR}, with signals captured by the RFSoC ADC, demonstrating the spectrometer’s capability to maintain precision across the tested frequency spectrum.}
    \label{fig:specNSDvsENOB}
\end{figure}

The consistent NSD and ENOB across the frequency range from 10 MHz to 100 MHz, as shown in Fig.~\ref{fig:specNSDvsENOB}, suggest that the PFB configuration with 62.5 kHz bin widths is effective in maintaining the necessary spectral resolution and minimizing quantization noise, critical for detecting the weak 21-cm signal. The system's ability to maintain linearity in NSD indicates that the chosen configuration supports a flat response and a sufficiently high SNR according to the ENOB metric, ensuring that the signal can be reliably extracted despite the noise power integrated across the wider measurement bandwidth. This linearity is essential for preserving the integrity of the 21-cm signal, allowing for precise characterization even under varied testing conditions.

ENOB across the frequency spectrum shows minimal deviation from the expected values of 12 bits despite slight variations due to signal distortion. The typical ENOB value of 11.5 bits confirms that lower distortion correlates with improved ENOB, which benefits from the normalization adjustments applied in the calculations. This adaptation helps mitigate the impact of low-frequency challenges, which are typically pronounced in radio astronomy. Although the Xilinx datasheet does not provide information on ENOB, the results match those of ADC converters used in radiometers. Characterization of the harmonics generated by the ADC remains significant for low-frequency radio astronomy.

Further analysis indicates that NSD is preferable to ENOB as a metric for assessing ADC's performance through SNR, which excludes harmonics. As the signal frequency nears the ADC's clock frequency, the ADC has limited time to accurately measure the signal's amplitude before the next sampling event, which compresses the dynamic range in the digital output. Additionally, the effectiveness of noise shaping and techniques dependent on oversampling or filtering across multiple samples diminishes, resulting in an increase in NSD and further distorting the signal representation. The measured NSD values around -152.5 dBFS/Hz are better than the typical -150 dBFS/Hz at 240 MHz reported by \cite{ZynqRFSoC2023} for similar conditions, this discrepancy likely stems from nonlinear effects introduced by the RF hardware, specifically the transformers used at balun board.

\subsubsection{Two-tone test}

From a mathematical standpoint, implementing the system’s transfer function within the framework of a Taylor series yields a spectrum of frequency products in the output. The two-tone present at the input of the transfer function yields output components with different frequency points and coefficients. The third-order products of (\(2\omega_1 - \omega_2\)) and (\(2\omega_2 - \omega_1\)), 3rd-order intermodulation distortion (IMD3), are considerably closer to the fundamental frequency. Consequently, attempts at their extraction can trigger a notable reduction in the SNR, effectively compromising the overall quality of the signal. In the Taylor series, their coefficients result in a 1 to 3 slope, meaning they amplify three times quicker than the fundamental signal with respect to input amplitude.

\begin{figure}
    \centering
    \includegraphics[width=\columnwidth, trim=0cm 1cm 0cm 2cm, clip]{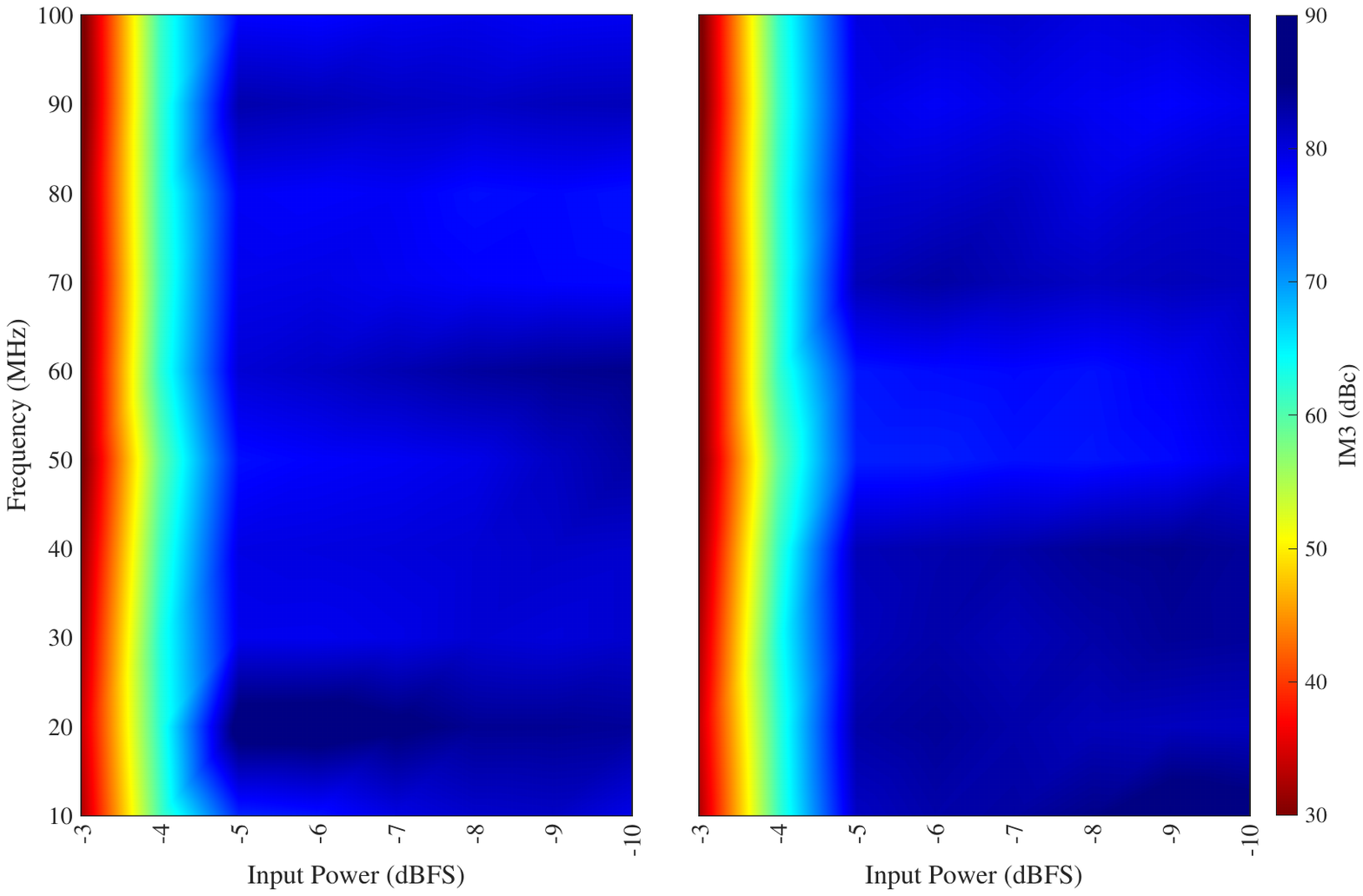}
    \caption{Comparison of IM3 lower (left) and upper (right) sidebands at different input power levels and frequencies, showing the performance consistency of a two-tone signal generator PNA-X.}
    \label{fig:IM3}
\end{figure}

The PNA-X was used to generate CW two tones with a spacing of 0.5 MHz. Fig.~\ref{fig:IM3} demonstrates the lower and upper IM3 values. It can be seen that around -5 dBFS, IM3 drops below -80 dBc. The data aligns with Xilinx datasheet values, indicating a -78 dBc performance of IM3 for the ADC at 240 MHz, with -7 dBFS input amplitude and 20 MHz spacing. 

When examining the upper and lower IM3 products in a nonlinear device, differences in their behavior can indicate the presence of memory effects, which refer to how the past behavior of the signal influences the current performance of the ADC. This often causes the device to behave differently based on its previous states, especially under varying signal conditions. Specifically, if the upper and lower IM3 products differ significantly, it might indicate asymmetrical memory effects. These effects can occur due to factors such as thermal lag, power supply variations, or component mismatches within the ADC that are affected by the envelope frequency. We analyzed these aspects using waterfall plots, where we observed that the left and right IM3 products are similar, and IM3 levels are high enough on both sides. We will investigate any potential impact of these minor differences in future studies.

\subsubsection{PFB channel leakage}

Channel leakage occurs when energy from a signal in a given frequency channel spreads into adjacent channels, potentially leading to significant distortions in the data collected. Even with narrow band filtering of PFB, the finite length of the FFT window and the properties of the window function itself can result in leakage. The 21-cm line is extremely faint and situated among various stronger astrophysical signals. Leakage across frequency channels, particularly the leakage of the various stronger signals, such as galactic synchrotron radiation or emissions from nearby galaxies, can lead to a misinterpretation of the weak 21-cm signal by either overwhelming it with noise from adjacent channels or by blending important spectral features.

\begin{figure}
    \centering
    \includegraphics[width=\columnwidth, trim=1cm 6cm 0.5cm 7cm, clip]{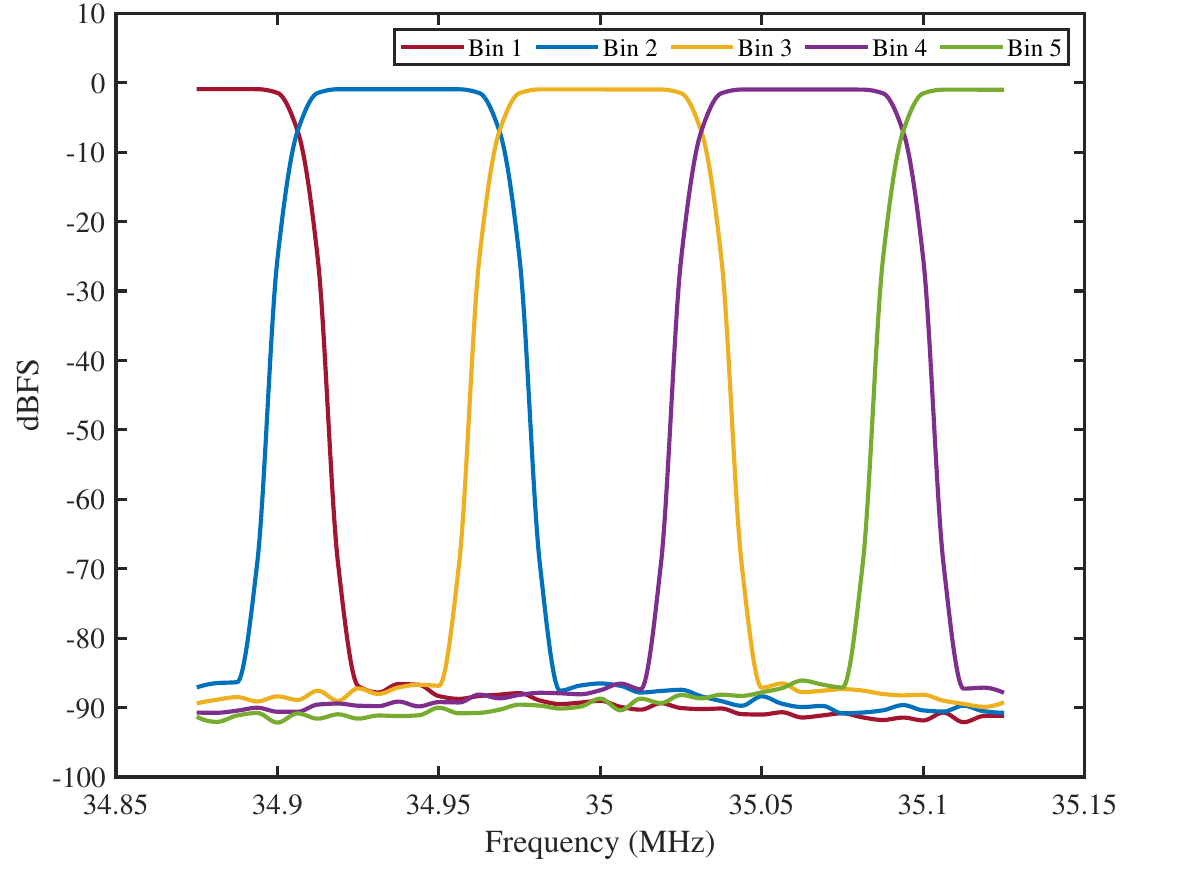}
    \caption{PFB channel isolation performance across five frequency bins around 35 MHz, showcasing the minimal leakage between channels. Signals were generated using a Keysight E8257D and captured by the RFSoC ADC.}
    \label{fig:channelLeakage}
\end{figure}

Sweeping signals at a frequency of 6.25 kHz were utilized to assess the channel leakage performance, as illustrated in Fig.~\ref{fig:channelLeakage}, of the PFB, which was configured for a channel width of 62.5 kHz for our spectrometer. While spectral leakage can be reduced, this requires more calculations in the FPGA, either by increasing sample numbers, reducing bin width, or implementing overlapping windows.

\subsection{Loopback test}

The previous experiment generated the signal using a signal generator to assess spectrometer performance. However, in CosmoCube, we will employ a calibration methodology that utilizes both DACs and ADCs within the RFSoC. Hence, we measured the loopback, where the coax cable connects the ADC of the RFSoC to the DAC of the same RFSoC by only employing AAF between the DAC and ADC connection. The losses between the frequency points were compensated. 

\begin{figure}
    \centering
    \includegraphics[width=\columnwidth, trim=0.2cm 6.54cm 0.5cm 6.5cm, clip]{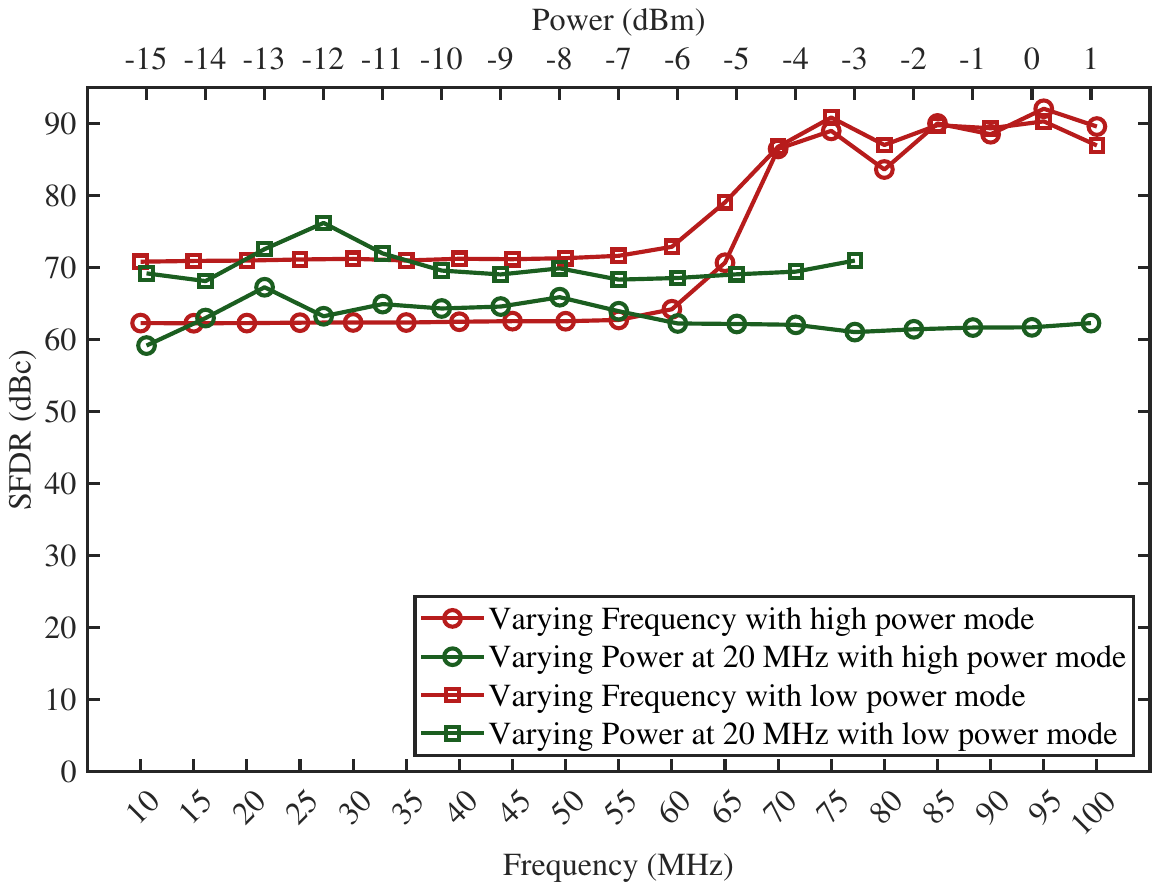}
    \caption{Performance comparison of SFDR across varying frequencies and power settings in RFSoC's loopback mode, highlighting high and low power modes at 20 MHz. A low-loss coaxial cable and an anti-aliasing filter (AAF) with a 105 MHz cutoff, designed using a 9th-order Chebyshev low-pass filter, are placed in between.}
    \label{fig:loopback}
\end{figure}

Fig.~\ref{fig:loopback} demonstrates the SFDR performance of the spectrometer by varying frequency and input power. Two modes, high and low power, of the same DAC, have been investigated. Our observations indicate that the SFDR remains consistent when either or both the 2nd and 3rd harmonics are within the measured bandwidth, which corresponds to frequency up to 60 MHz. In scenarios where only folded frequency components are present within the bandwidth, both low and high-power modes exhibited improved performance, yet the SFDR was comparable between the two modes. Although the significant shift in SFDR from 60 MHz to 70 MHz necessitates adjustments to the calibration algorithm, the generally stable performance of the SFDR across two bands, low and high frequency, suggests that managing the calibration should be straightforward.

Additionally, varying the input power at 20 MHz confirmed that the dynamic range remains stable in both high and low power modes. The low-power mode consistently maintains a narrower SFDR spread, which suggests that it is more resilient to spectral anomalies or harmonic distortions under varied testing conditions. The robustness of the SFDR across the entire bandwidth significantly assists in the calibration process by ensuring consistent measurement accuracy and reliability.

\subsection{Noise measurements with varying temperatures and integration time}

A lower noise floor enhances sensitivity to the extremely faint 21-cm signal, which is crucial for distinguishing them from background noise. Accurate noise floor characterization improves data reliability, enabling more precise insights into the Universe’s earliest structures.

We used a thermal chamber to test the stability of the ZCU111 RFSoC under various temperature conditions, adhering to the device's specified operating range of 0$^\circ$C to 45$^\circ$C, as indicated in its datasheet. Although thermal systems will be in place to ensure temperature stability by heating and cooling the receiver part of the system inside the CubeSat, we used a thermal chamber to verify that the spectrometer could reliably maintain its performance and accuracy in the extreme thermal environments encountered in space. By systematically adjusting the temperature within the chamber, we observed the device's response to thermal stress, ensuring that it could handle the fluctuations without degrading its functionality.

\begin{figure}
    \centering
    \includegraphics[width=0.9\columnwidth, trim=1cm 6.5cm 1cm 7cm, clip]{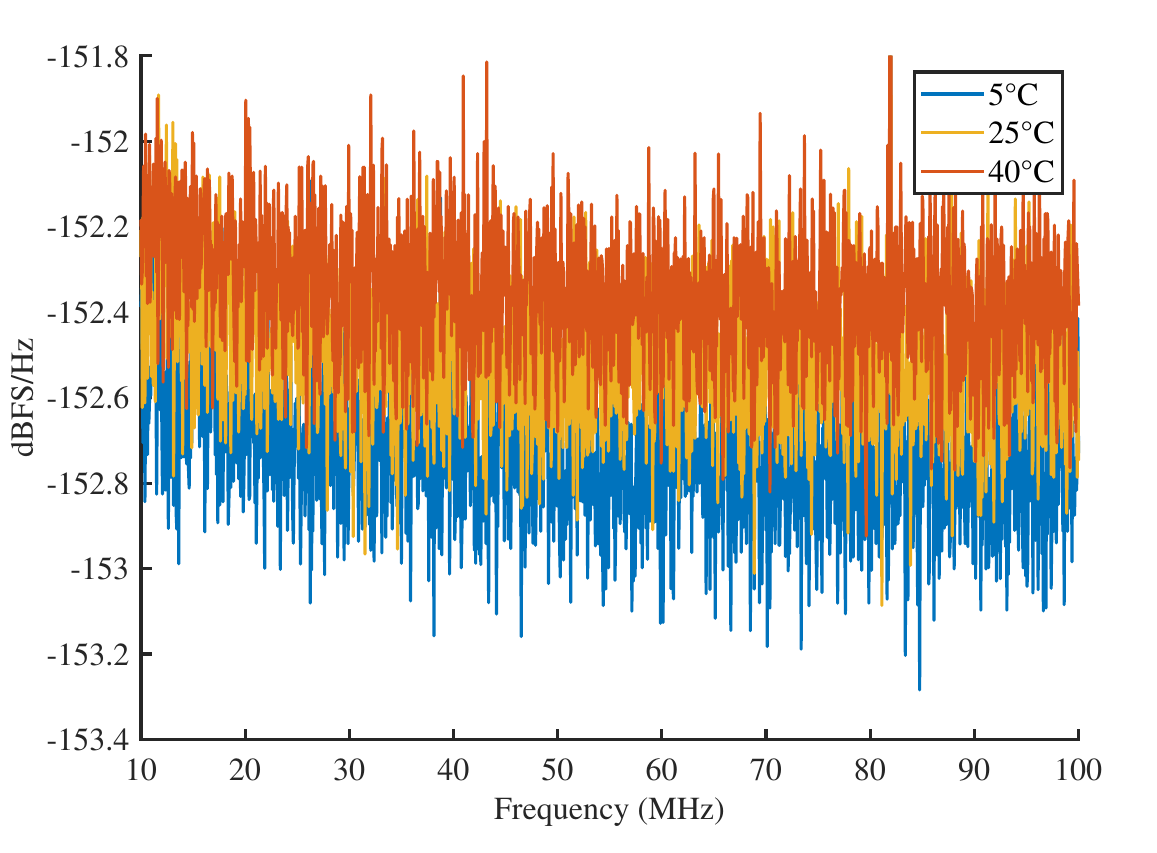}
    \caption{Noise floor measurements for a terminated 50 ohm load at various ambient temperatures. The noise floor was recorded at ambient temperatures of 5$^\circ$C (blue line), 25$^\circ$C (orange line), and 40$^\circ$C (red line), and measured heatsink temperatures were at 24.4$^\circ$C, 51.6$^\circ$C, and 69.8$^\circ$C with corresponding FPGA junction temperatures of 24.2$^\circ$C, 53$^\circ$C, and 71.4$^\circ$C respectively. Each measurement ran for 30 minutes in order to ensure temperature stability before taking the measurements. Integration time for each measurement was 2 ms.}
    \label{fig:NoiseTemp}
\end{figure}

We recorded a noise floor of approximately -152.5 dBFS/Hz at an environmental temperature of 25$^\circ$C, as shown in Fig.~\ref{fig:NoiseTemp}. The datasheet for the ZCU111 specifies a typical noise floor of about -152 dBFS/Hz at a frequency of 240 MHz. The observed $\pm$0.2 dB variation under extreme temperature conditions confirms the spectrometer's stable operation, demonstrating its robustness and reliability for precise measurements in the challenging thermal environment of space. However, this stability must also be evaluated in light of the temperature gradients that occur as the satellite transitions rapidly between sunlight and the Earth's shadow, potentially imposing additional thermal stress on the system. Nevertheless, since observations will primarily occur during occultation, where thermal variation is expected to be low and manageable by the payload's thermal management system, these effects should be minimized. Furthermore, this thermal variation will be recorded and considered during the data analysis.

\begin{figure}
    \centering
    \includegraphics[width=\columnwidth, trim=1cm 6cm 0.5cm 7cm, clip]{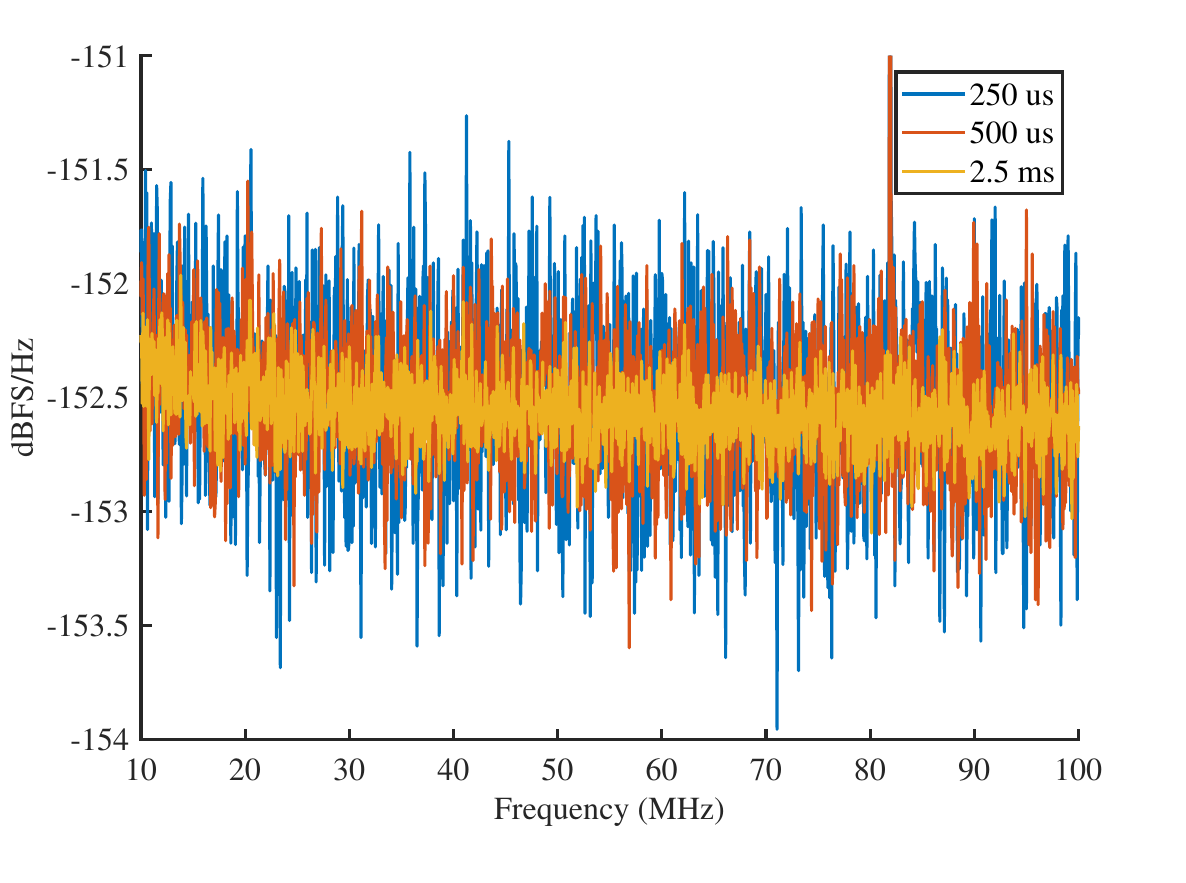}
    \caption{Noise floor measurements with varying integration time while ADC is terminated by a 50 ohm load. The heatsink temperature was set at around 53$^\circ$C and the junction temperature at around 55$^\circ$C, while the ambient temperature in the chamber was maintained at around 25$^\circ$C.}
    \label{fig:NoiseIntTime}
\end{figure}

The integration time is calculated by multiplying the number of samples, 2048, acquired by the ADC by the number of averaged frames of 250, 500, and 2500 and then dividing the result by the sampling rate of 2048 MSPS. Fig.~\ref{fig:NoiseIntTime} illustrates how the noise floor varies with different integration times, demonstrating the impact of integration duration on the stability and clarity of the signal detected by the spectrometer.

The noise floor of the same ADC was measured on two separate occasions, showing a variation of approximately $\pm0.015$ dB. This is comparable to the $\pm0.025$ dB variations observed by \cite{liu2020} in their analysis of the ZCU111 for wideband radiometer applications. The variation between these two works might be evidence of temperature stability, as the board's temperature in this work was maintained in a thermal chamber for some time to ensure stability.

\section{Conclusions}

In conclusion, the CosmoCube instrument represents an advancement in lunar and space-based radio astronomy, particularly in exploring the faint 21-cm hydrogen line from the Dark Ages. By overcoming terrestrial limitations and harnessing innovative technologies such as the Xilinx RFSoC, which, while highly advanced, is not yet fully space-qualified but will undergo testing and can be placed in a tailored space-qualified hermetically sealed package, CosmoCube provides unprecedented sensitivity and resolution, offering new insights into the early Universe's structure and evolution. Unlike the custom-built analog components in systems like EDGES, REACH, and SARAS, the RFSoC offers a higher level of integration and availability of various sources in the firmware development, which is advantageous for the rapid and cost-effective development of the spectrometer. However, it requires careful consideration of its space-readiness, particularly in terms of radiation hardness and thermal management. Utilizing a channelizer in the spectrometer, we achieved an ENOB of approximately 11.5 and an NSD of around -152.5 dBFS/Hz, highlighting the system’s high precision and low noise performance across the frequency range of interest. Even with the exceptional performance showing only a $\pm$0.2 dB variation under extreme temperatures, thermal systems are still required to maintain stability, particularly for the calibration of the system. However, it is important to note that no receiver has yet been coupled to the system, which could introduce new challenges or alter its performance. The integration of the receiver may reveal additional complexities, requiring further adjustments to optimize signal integrity and overall system behavior. Furthermore, the need for stringent filtering techniques was evident, as harmonic levels impacted data integrity. The implementation of specialized filters and characterization of these spurious signals will be critical in future refinements to reduce signal contamination and enhance the overall quality of the cosmological data collected by CosmoCube. Further analysis of power consumption will be made to optimize the balance between system performance and operational longevity, ensuring the mission’s sustainability and efficiency in the harsh conditions of space.

\section*{Acknowledgements}

The authors would like to thank Dominic Anstey and Harry T. J. Bevins for their work with the Bayesian framework, William Grainger for thermal analysis, and David Bacon for lunar orbit analysis. We acknowledge the support by the UK Space Agency (UKSA) with the funding codes UKSA ST/Z000548/1 and UKSA ST/Z000556/1. Kaan Artuc thanks RF-Lambda for the funding. Eloy de Lera Acedo acknowledges the Science and Technology Facilities Council (STFC) for their funding support through an STFC Ernest Rutherford Fellowship.

\section*{Data Availability}
 
The data underlying this article will be shared on reasonable request to the corresponding author.

\bibliographystyle{rasti}
\bibliography{rasti_template} 



\bsp	
\label{lastpage}
\end{document}